\documentclass[sigconf,authorversion,nonacm]{acmart}

\usepackage{url}
\usepackage{amsmath,amsfonts,multirow}
\usepackage{algorithmic}
\usepackage{graphicx,booktabs}
\usepackage{textcomp}
\usepackage{xcolor} 
\usepackage{wasysym}

\usepackage{verbatim}
\usepackage[]{algorithm2e}
\usepackage{ifthen}
\usepackage[normalem]{ulem}
\useunder{\uline}{\ul}{}
\usepackage{pifont}
\usepackage{amsmath}
\usepackage{csquotes}
\usepackage{enumitem} %
\usepackage{listings} %
\usepackage{ulem}

\usepackage[colorinlistoftodos,textwidth=1.22cm]{todonotes}

\newcommand{\review}[1]{\todo[color=gray!30]{#1}{}}

\renewcommand\review[1]{} %

\newcommand{\market}{\texttt{ImpaaS.ru}}
\newcommand{\hackforum}{\texttt{Torum}} %
\newcommand{\cardingmarket}{\texttt{Crdclub}} %
\newcommand{\impaas}{\texttt{IMPaaS}}

\newcommand{\resources}{\texttt{Resources}}
\newcommand{\fingerprints}{\texttt{Fingerprints}}
\newcommand{\cookies}{\texttt{Cookies}}

\newcommand{\crypto}{\texttt{Cryptocurrency}}
\newcommand{\money}{\texttt{Money transfer}}
\newcommand{\commerce}{\texttt{Commerce}}
\newcommand{\social}{\texttt{Social}}
\newcommand{\services}{\texttt{Services}}
\newcommand{\other}{\texttt{Other}}

\def\BibTeX{{\rm B\kern-.05em{\sc i\kern-.025em b}\kern-.08em
    T\kern-.1667em\lower.7ex\hbox{E}\kern-.125emX}}

\begin{document}
\fancyhead{}

\title{\texttt{Impersonation-as-a-Service}: Characterizing the Emerging Criminal Infrastructure for User Impersonation at Scale}%

\author{Michele Campobasso}
\email{m.campobasso@tue.nl}
\affiliation{%
  \institution{Eindhoven University of Technology}
  \city{Eindhoven}
  \country{Netherlands}
}

\author{Luca Allodi}
\email{l.allodi@tue.nl}
\affiliation{%
  \institution{Eindhoven University of Technology}
  \city{Eindhoven}
  \country{Netherlands}
}

\begin{abstract}
In this paper we provide evidence of an emerging criminal infrastructure enabling impersonation attacks at scale. \textit{Impersonation-as-a-Service} (\impaas) allows attackers to systematically collect and enforce user profiles (consisting of user credentials, cookies, device and behavioural fingerprints, and other metadata) to circumvent risk-based authentication system and effectively bypass multi-factor authentication mechanisms. We present the \impaas\ model and evaluate its implementation by analysing the operation of a large, invite-only, Russian \impaas\ platform providing user profiles for more than $260'000$ Internet users worldwide. Our findings suggest that the \impaas\ model is growing, and provides the mechanisms needed to systematically evade authentication controls across multiple platforms, while providing attackers with a reliable, up-to-date, and semi-automated environment enabling target selection and user impersonation against Internet users as scale.
\end{abstract}

\keywords{user profiling; impersonation attacks; impersonation-as-a-service; threat modeling}
\maketitle

\section{Introduction}

In recent years there has been a surge in criminal infrastructures supporting cyberattacks and cybercrime activities at large~\cite{Grier-12-CCS,allodi2017economic,Caballero-2011-USS}. For example, \textit{exploitation-as-a-service} and \textit{pay-per-install} provide a set of attack technologies generally aimed at infecting systems or controlling bots that are then employed to launch, for example, DDoS attacks, or subsequent malware and phishing campaigns (e.g., to harvest credit card numbers or steal credentials).
An important problem in any venture, let alone a criminal one, is the ability to \textit{systematically} monetize the effort that goes into it~\cite{Herley-2009-NSPW}. In criminal enterprises, monetization is not necessarily an easy feat: whereas re-selling or giving access to infected systems to fellow criminals alleviates the problem for whom generates the infection (e.g., the \textit{bot herder}~\cite{holt2016examining,binsalleeh2010analysis}), the problem of assigning a price to each bot remains~\cite{anderson2006economics}. \review{R3.6} 
Whereas the dynamics of demand and offer in the underground are likely to play a role in this setting (and remain an important open question to investigate in this domain), another key factor in determining the value of an infected system is the information it manages and/or processes; %
for example, access to the email account(s) of an Internet user may have a different value, to attackers, than access to a user profile with a server-stored credit card number (e.g., an e-commerce website). On the other hand, it is not yet clear how (and if) attackers can \textit{systematically} employ those credentials to impersonate Internet users at large, particularly in the presence of multi-factor authentication systems whereby a username and password alone are not sufficient to gain access to an Internet account. 

Credential theft and re-selling in underground communities have been studied multiple times in the literature; 
for example, recent studies provide an in-depth view of what happens to credentials after they have been stolen~\cite{Onaolapo-IMC-16}, and their employment for final attacks~\cite{thomas2017data}. Similarly,  several studies investigate the attack vectors that allow attackers to obtain the credentials in the first place, ranging from (targeted) phishing and phishing kits, to malware infections at scale~\cite{bonneau2015passwords,bursztein2014handcrafted,Onaolapo-IMC-16}. 
On the other hand, a systematic employment of the stolen credentials remains out of reach for most attackers: credentials stolen from the underground may be accessed by multiple criminals, effectively destroying their value for later accesses~\cite{Herley-2009-NSPW}; similarly, the effort required to monetize access to stolen or hijacked user accounts does not scale well with the number of available accounts~\cite{Herley-2009-NSPW,Herley-WEIS-12}.
In particular, protection systems such as multi-factor and risk-based authentication systems severely limit the capabilities of attackers to effectively employ stolen credentials, requiring the employment to more sophisticated attack vectors than a simple credentials dump~\cite{thomas2017data}. Risk-based authentication systems receive user authentication requests and are responsible to decide whether additional multi-factor authentication is required for that session, or if the provided (valid) password suffices to grant access to the user requesting it. The idea behind risk-based authentication is that, by `measuring' certain characteristics of the user environment (i.e., its fingerprint~\cite{alaca2016device}), the authenticating system can build a `risk profile' associated to that request as a function of the distance between the current fingerprint and the profile associated to the requesting user. If the mismatch is too large, the risk-based authentication system will defer the decision to a multi-factor mechanism (e.g., requesting a code sent to a trusted device or account, such as a mobile phone or an email account); on the other hand, if no anomaly in the user profile is detected, the risk-based authentication system will -- in most cases -- grant access just with the password.

This mechanism is a significant obstacle to a successful impersonation attack, as the very high dimensionality of a user fingerprint makes it impossible, for an attacker, to \textit{systematically} reproduce it for arbitrary users from scratch~\cite{thomas2017data,alaca2016device}. A recent study by Thomas et al.~\cite{thomas2017data} highlights how modern phishing kits~\cite{oest2018inside} are equipped with fingerprinting modules that, together with the user credentials, obtain a measurement of the user's environment that can be re-used to circumvent risk-based systems. On the other hand, obtaining these user profiles require systematic efforts to phish targets, perhaps across different platforms, and may not provide reliable and stable measures of a user's fingerprint as the victim's interaction with the attacker website may not accurately reflect the victim's interaction with the legitimate website (e.g., for behavioural fingerprinting~\cite{stringhini2015ain,chen2009large}). Overall, traditional attack strategies seem unsuitable to reliably obtain, update, and enforce user profiles.

In this paper we provide evidence of a new emerging criminal infrastructure for \textit{Impersonation-as-a-Service}, that relies on custom malware and a marketplace platform to systematize the delivery of complete \textit{user profiles} to attackers. A user profile on an \impaas\ service comes complete with stolen credentials for multiple platforms, the ability to either reproduce or re-generate a user fingerprint from the stolen data, and a software bundle to enforce the user profile during an authentication session. To study the presence of the \impaas\ model in the wild, we provide an in-depth analysis of a large criminal platform (\market) providing, at the time of writing, more than $260'000$ profiles of Internet users, globally. \market\ is an emerging, invite-only, Russian \impaas\ platform currently operating in the underground. To evaluate the nature of \impaas\ operations, we dissect the process behind the \textit{acquisition, selection, and enforcement} of stolen user profiles enabled by the \impaas\ model, and provide a detailed evaluation of the characteristics of \market, its extension, the characteristics of the user profiles it provides to final attackers, and the relative effect of different user profile characteristics on its value.

\textbf{Scope and contribution.} The contribution of this paper is three-fold:
\begin{itemize}
    \item we provide the first characterization of the \impaas\ model for the systematization of impersonation attacks at scale;
    \item we provide an evaluation of a large, invite-only, emergent Russian \impaas\ platform that automates the collection, provision and enforcement of user profiles collected worldwide;
    \item we provide insights on the relative effects of different user profile characteristics on the value of the user profile, and quantify these effects.
\end{itemize}
A detailed technical analysis of the malware for the user profile exfiltration and enforcement is out of the scope of the present paper.

This paper proceeds as follows: Section~\ref{sec:backgroun} set the background for impersonation attacks and their relation to existent countermeasures; Section~\ref{sec:impaas} introduces the \impaas\ model for impersonation attacks at scale, and Section~\ref{sec:characterizing} describes the \market\ marketplace implementing it, and our infiltration and data collection strategy. \market\ operations are analysed in Section~\ref{sec:dataanalysis}. Section~\ref{sec:discussion} discusses our findings, and Section~\ref{sec:conclusion} concludes the paper.

\section{Background and related work}
\label{sec:backgroun}

\subsection{User impersonation attacks}

With the rise of sophisticated web applications, much of a user's Internet activity happens by accessing a multitude of remote services, from banking to e-commerce and social network platforms, through the browser. Most of these services will have authentication mechanisms that are meant to grant access to the underlying service to the authorized user(s) only. From an attacker's perspective, user impersonation provides a large portfolio of additional attack opportunities, ranging from economic gain~\cite{allodi2017economic,Franklin-2007-CCS} to more targeted scenarios such as targeted-phishing~\cite{Ho-USENIX-19} and violent crimes~\cite{Havron-USENIX-19}.

Password-based authentication (PBA) is the most common (first) barrier attackers have to overcome to perform an impersonation attack. Whereas passwords have proven difficult to securely handle, are prone to leaks and to off-line attacks~\cite{morris1979password, yan2012limitations} and still present severe usability problems~\cite{stobert2014agony}, they represent the most widespread means of authentication online~\cite{bonneau2012quest,bonneau2015passwords}. PBA requires users to create a non trivial secret, not to reuse it across several services and to memorize both the secret and where it has been used; nonetheless, several studies indicate that up to $\approx 90\%$ of users reuse passwords or small variations thereof across several services~\cite{das2014tangled,ion2015no}. 

Whereas this leaves room for password guessing attack, additional attack vectors (such as malware and phishing~\cite{thomas2017data,bursztein2014handcrafted}) can be used to obtain user passwords, regardless of their complexity. %
In general, hijacked accounts can allow adversaries to tap into social connections of victims to compromise additional accounts~\cite{thomas2014consequences,gao2010detecting}, by creating targeted social-engineering attacks against their circle of trust or by spamming malicious content~\cite{sabillon2016cybercriminals}, liquidate financial assets~\cite{ioactive_2012}, steal sensitive information with the aim of blackmailing users~\cite{bursztein2014handcrafted,sabillon2016cybercriminals} and sextortion~\cite{wittes2016sextortion}. 
Additionally, stolen user credentials are oftentimes made available to the cybercrime community through underground markets~\cite{Onaolapo-IMC-16,thomas2017data}. These markets generally provide `dumps' of stolen credentials obtained from data leaks from an affected platform, or as a result of an extensive phishing campaign targeting its users~\cite{thomas2017data}; common target platforms include banking or trading websites, cryptocurrency services, pornographic websites, and other internet services. A recent estimation calculates that, between March 2016 and March 2017, 1.9 billion phished credentials has been sold through the underground markets~\cite{thomas2017data}.

\subsection{Countermeasures to attacks against PBA}

\textbf{Multi-Factor Authentication.} To mitigate the shortcomings of authentication mechanisms relying solely on passwords, web platforms have started adopting additional authentication measures such as Multi-Factor Authentication (MFA). MFA moved the authentication paradigm from (solely) something that the user \textit{knows} (e.g. a password) to something the user \textit{has} (e.g., a token)~\cite{thomas2017data,Dmitrienko-FCDS-14}. 
This is achieved mainly with a combination of a pair of valid credentials and a One Time Passcode (OTP) received via some trusted component such as a mobile phone, email, or a hardware token~\cite{Dmitrienko-FCDS-14}. Albeit possible attack scenarios exist where the attacker can obtain the information required for the authentication almost in real-time (stolen token generator, compromised email, SIM swap attacks~\cite{mulliner2013sms}, etc.), MFA dramatically increases the costs for an attacker, and is widely regarded as an effective countermeasure to password-based impersonation attacks~\cite{thomas2017data}. Nonetheless, MFA is not devoid of security problems, perhaps most notably related to its usability~\cite{208153}, concerns on token-recovery mechanisms, and third-party trust~\cite{bonneau2012quest}. %

\smallskip
\noindent\textbf{Risk-Based Authentication.} Partly to mitigate the usability problem, \textit{Risk-Based Authentication} (RBA) is oftentimes adopted as a means to evaluate whether the authenticating user is (likely to be) the one that has, historically, access to a specific account. %
RBA is an adaptive security technique aiming to strengthen password-based authentication by monitoring how unexpected  or suspicious a login attempt is from the perspective of the authenticating service~\cite{wiefling2019really, 208153, thomas2017data}. During the authentication, the RBA system monitors both behavioral and technical characteristics of the user and of the device, producing a \textit{fingerprint} of the authenticating user~\cite{wiefling2019really}. RBA computes a risk score associated to the ongoing authentication by comparing the existent profile of the authenticating user against the features collected for that instance of the authentication.
The features vary from basic information such as User-Agent, system time and OS, to environmental or behavioral features, such as system language, keyboard layout, fonts and plugins installed, mouse movement, geolocation and keystroke speed~\cite{alaca2016device, wiefling2019really, freeman2016you, thomas2017data}. Whereas the high dimensionality of this data generates, with high probability, \textit{unique} `fingerprints' of a user, these are not necessarily \textit{stable} in time (as, for example, users may access the service from multiple or new systems, may update software configurations, or authenticate from different locations).
Depending on the computed risk score for that transaction, the authenticating service may grant access to the user with only a valid password (if the risk level is low), or require additional authentication factors (e.g., codes sent to associated email accounts, SMS verification) or even deny access for higher risk levels~\cite{wiefling2019really, 208153}. This mechanism relies on the assumption that attackers cannot systematically re-create the profile of the victim, unless the attacker is already in control of a user's system.

Following the implementation of RBA techniques across critical services, adversaries developed sophisticated solutions aiming to impersonate the user profile of the authenticating user. Recent literature has shown that phishing kits have developed capabilities to obtain user profiles that can then be re-used by the attacker; similarly, recent malware has been specifically engineered to report user activity back to the attacker~\cite{thomas2017data}. In particular, Thomas et al.~\cite{thomas2017data} highlight the improved capabilities of phishing kits in collecting information related to victims, including geographical location, browser metadata and answers to security questions; they found that attacks relying on user profile information collected from phishing kits are 40 times more likely to be successful than `regular' attacks based on leaked credentials. On the other hand, the collection of user profile information does not scale well across users and platforms as user profiles may vary with time, across services, and must to be collected by the attacker through additional attack means (e.g., phishing).

\subsection{Analysis of current attack strategies} 

\noindent\textbf{Attack capabilities.} From the analysis above, we identify six capabilities required to systematically bypass RBA systems.

\smallskip
\noindent\textit{Password authentication.} At the very minimum, an attacker needs the authentication credentials of the victim.

\smallskip
\noindent\textit{User profiling.} To attempt circumventing RBA systems, an attacker should have an accurate measurement of the victim's profile/fingerprint for that platform.

\smallskip
\noindent\textit{Multi-platform.} The attacker may need to access multiple web platforms to bypass some MFA controls (e.g. tokens or OTPs sent to an email account of the victim). Authentication credentials and user profiles need to be collected for these additional platforms as well. The capability of impersonating the victim on multiple platforms further increases the attack surface in scope of the attacker.

\smallskip
\noindent\textit{Profile updates.} User profiles are unique but not necessarily stable. For example, a user may update a password, change software configuration, or access the service from a different geographical region. These changes may invalidate previously collected profiles for that user, which may therefore require updating.

\smallskip
\noindent\textit{Infection infrastructure.} The attacker requires an infrastructure to infect users, and collect and update the collected user profiles. This has to be maintained as defensive capabilities evolve (e.g. blacklisting of an employed phishing domain), and may require the acquisition of external services (e.g., for an infection update~\cite{Grier-12-CCS,Caballero-2011-USS}).

\smallskip
\noindent\textit{Automated profile enforcement.} Once a profile is collected, the attacker needs to enforce it when authenticating on the platform. Whereas some aspects of the profile are easy to reproduce (e.g., user agent, screen resolution), others are not (e.g., installed fonts/plugins, keystroke speed, mouse movements, etc.). As profiles change across users and platforms, the attacker likely needs a system capable of enforcing the collected profiles in an automated fashion.

\smallskip
\noindent\textbf{Analysis across attack strategies.} Kurt et al.~\cite{thomas2017data} identify three main strategies for impersonation attacks. Table~\ref{tab:rbaevasion} provides an overview of their capabilities.

\smallskip
\noindent\textit{Leaked credentials.} credentials derived from data breaches on a platform. Leaked credentials are generally traded in bulk in underground forums; the leaked data oftentimes only contain associations between usernames and (hashed) password, with no user profile information. The data is static and if a user changes the password, the information owned by the attacker loses all value. As the leak concerns only one platform (and multiple leaks are likely unrelated to each other), cross-platform attacks against one user are not enabled by this attack strategy. However, password-reuse attacks may provide the attacker with access to additional platforms on top of the one that suffered the leak.

\smallskip
\noindent\textit{Phishing kits.} attackers can employ kits to deploy phishing websites aimed at stealing user credentials. As users directly interact with the phishing kit, user profiling can be achieved by injecting fingerprinting code in the phishing webpage~\cite{thomas2017data}. The profiles derived through phishing kits are however limited to only one occurrence of the authentication (on the phishing website) and may be incomplete or inaccurate. For example, the employment of password manager software may hinder the realism of the derived fingerprint (e.g., in terms of input time or user behaviour on the page) when compared to the one measured by the original platform. To achieve multi-platform capabilities, the attacker must develop or acquire a phishing kit for each of the phished platforms, and collect the relevant data through separate attacks against the same user.

\smallskip
\noindent\textit{Malware.} the attacker has access to the system through a keylogger or trojan/bot. This requires the attacker to either purchase/rent the infected system~\cite{Grier-12-CCS}, or create the infection themselves (e.g., through malware attached to a phishing email, or through Pay-per-Install services~\cite{Caballero-2011-USS}). 
Due to the specificity of the attack, custom malware is likely needed to collect and update the profiles. As the attacker is virtually already in full control of the user system, they can collect user profiles related to any platform accessed by the victim. However, due to the position of the attacker, most of the impact (e.g., email access or web session hijacking) can be achieved through malware without the need of collecting the user profiles to then replicate them at a later stage. 

\begin{table}[t]
    \centering
    \caption{Overview of impersonation attack capabilities. %
    }
    \vspace{-0.08in}
    \begin{minipage}{0.95\columnwidth}
    \footnotesize
    \CIRCLE\ indicates full systematic capability; \RIGHTcircle\ indicates systematic capability only after specific engineering effort from attacker; \Circle\ indicates no systematic capability.
    \smallskip
    \end{minipage}
    \label{tab:rbaevasion}
    \begin{tabular}{lp{0.05\columnwidth}p{0.13\columnwidth}p{0.12\columnwidth}p{0.12\columnwidth}}
    \toprule
        & Leak & Phishing kits & Malware & ImpaaS \\
        \midrule
         Password auth. & \RIGHTcircle & \CIRCLE &\CIRCLE &\CIRCLE \\
         User profiling & \Circle\ & \CIRCLE & \RIGHTcircle & \CIRCLE \\
         Multi-platform & \Circle\ & \RIGHTcircle& \RIGHTcircle& \CIRCLE \\
         Profile updates & \Circle\ & \Circle\ & \CIRCLE & \CIRCLE\\
         Infection infrastructure & \Circle\ & \Circle\ & \RIGHTcircle\ & \CIRCLE\\
         Automated profile enf. & \Circle\ & \Circle\ & \Circle\ & \CIRCLE\\
         \bottomrule
    \end{tabular}
\end{table}

\begin{figure*}[t]
\centering
\includegraphics[width=0.99\textwidth]{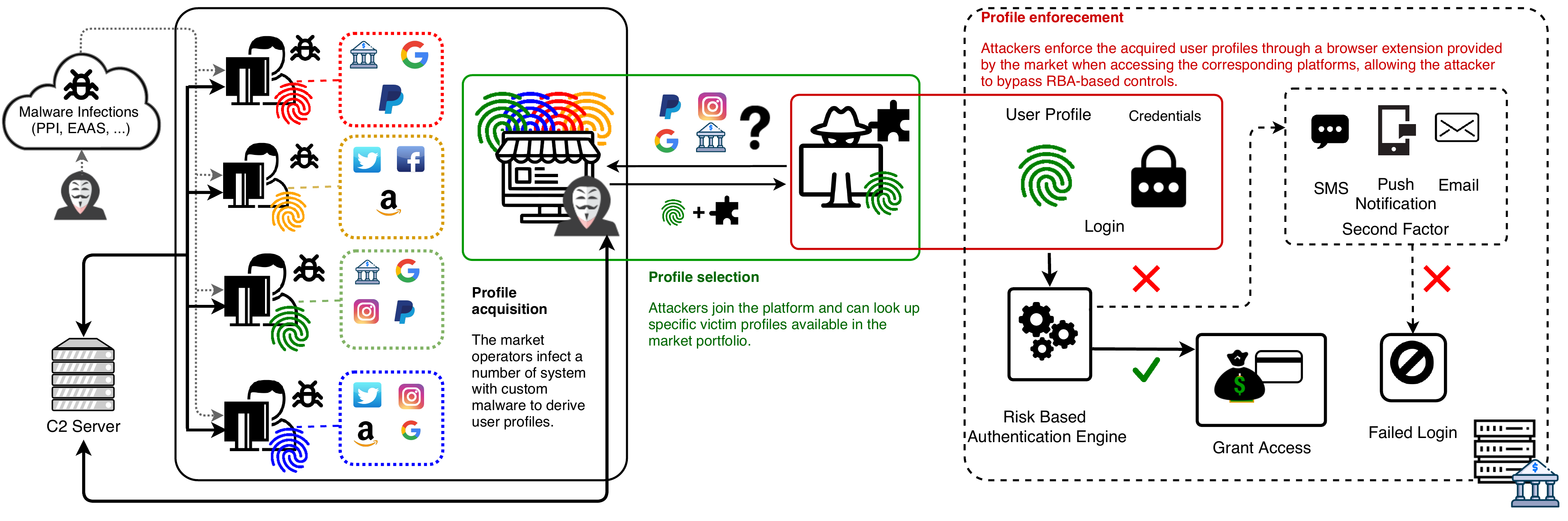}
\begin{minipage}{0.98\textwidth}
\footnotesize
By only owning the credentials of the victim, the attacker cannot bypass the MFA as the Risk-Based Authentication (RBA) system will detect an anomaly in the profile of the authenticating user. By relying on \textit{Impersonation-as-a-Service} (\impaas), the attacker can reliably impersonate that profile by providing the values the RBA system expects for that user. \impaas\ obtains user profiles from a (large) botnet, and provides them in \textit{bundles} as user profiles. An attacker purchases a user's profile(s) on the \impaas\ platform together with a browser extension that, provided the victim's profile as an input, reproduces it when accessing a service. %
\end{minipage}
\caption{Diagram of Impersonation-as-a-Service operations.}

\label{fig:fingerprinting}
\end{figure*}

\section{The Impersonation-as-a-Service model}
\label{sec:impaas}

In this paper we describe evidence of a new emerging attack model, namely \textit{Impersonation-as-a-Service} (\impaas\ for short), and the criminal infrastructure supporting it. 

\impaas\ directly addresses the main limitations of the `traditional' impersonation attack strategies highlighted above by moving the acquisition and enforcement of victim profiles from an \textit{ad-hoc} process to a \textit{systematic} one. An overview of the comparison between \impaas\ and current vectors for impersonation attacks is summarized in Table~\ref{tab:rbaevasion}.  Figure~\ref{fig:fingerprinting} provides a birds' eye view of the attack process, from profile \textit{acquisition}, to \textit{selection} and \textit{enforcement}. \impaas\ operators rely on widespread malware infections to acquire `user profiles' globally, and provide these profiles as `goods' via the underground economy through a dedicated marketplace. As a result, attackers can acquire systematic access to a large set of user profiles spanning multiple platforms (social networks, email, corporate accounts, banking/cryptocurrency, etc.), alongside associated credentials and cookies; attackers can select the profiles they are most interested in based on a number of features, including the geographic location linked to the profile, the platforms for which impersonation data is available, amount of stolen cookies, date of profile acquisition, and others. The user profiles available on the \impaas\ platform are automatically updated by the underlying infrastructure (e.g., as users change software configuration, or update passwords); %
further, the attacker can easily enforce and switch across the acquired user profiles by means of a dedicated browser extension provided by the \impaas\ platform, effectively commodifying the systematic impersonation of Internet users at large across multiple platforms. 

\paragraph{Profile acquisition} The \impaas\ infrastructure is fueled by a botnet whose goal is, rather than solely collecting credit card information or banking credentials, to provide the information needed to replicate the user profiles of the infected victims across the online platforms on which affected users are active.
The malware distribution is independent from the \impaas\ model: it can be delivered through phishing campaigns, targeted attacks, pay-per-install~\cite{Caballero-2011-USS} or exploitation-as-a-service infrastructures~\cite{Grier-12-CCS}. Through the chosen attack vector, the attacker installs on the victim system custom malware engineered to collect user credentials and cookies from the victim's browsers; the custom malware further collects a large set of technical and (user) behavioral information that can be replicated, by means of the infrastructure itself, to fully emulate the user; these include the fingerprint(s) of the victim's browser(s) and other behavioral metadata that uniquely identify the user, such as network activity, browser history, cookie data, and interactions with the user interface of the platform. As profiles are fetched by means of a persistent malware infection, the infrastructure can provide updates of the profile data and credentials for each affected user. The harvested profiles and the respective updates are then pushed to the \impaas\ servers.

\paragraph{Profile selection}
An \impaas\ operator provides the harvested user profiles to interested attackers via a dedicated marketplace. The marketplace provides an overview of the characteristics of the collected profiles available for purchase, such that the attacker can select which profiles best fit their goal by searching for victim profiles that show specific features, such as a certain geographic location, web services for which stolen credentials are available, presence of cookies, etc. 
Albeit less targeted than allowed by a spear-phishing attack scenario, the selection procedure allows for a high degree of precision on the characteristics and/or environment of the user. For example, by browsing though the available credentials it is possible to identify users operating in a specific environment (e.g. a specific corporation, university, or other organizations), or with profiles on platforms of interest to the attacker. %
Once an attacker has identified their victim(s), the attacker can then proceed to buy the selected profiles. This can be achieved through the usual payment methods adopted in the cybercrime markets, such as via cryptocurrency payments to the marketplace, and/or by relaying the payment through a third-party escrow service. Importantly, as each profile can be purchased individually, the \impaas\ platform is in the position of removing purchased profiles from the marketplace listings, thus potentially reassuring the customer that they are the only one (next to the platform operators) with access to that profile.

\paragraph{Profile enforcement}
The \impaas\ platform provides their customers with a customized software bundle that includes a custom browser (based on open-source projects) and a browser extension that allows attackers to fetch and `enforce' the purchased user profiles during the attacker's browsing session on that platform. Based on the profiles selected and purchased by the attacker, the software provided by the \impaas\ platform recreates a browsing environment that replicates the victim's environment by instantiating exact copies of the stolen cookies and user credentials, and by spoofing other information on the victims' systems (e.g., installed fonts/plugins, browser agent, \ldots). Further, the profile enforcement system provides cookies that embed behavioral metadata derived from the victim~\cite{chen2009large} without requiring explicit action from the attacker, and provides SOCKS5 proxy solutions to spoof the usual geographic location of the victim.

\section{Characterizing ImpaaS in the wild}
\label{sec:characterizing}

In this section, we describe the operations of an emergent, invite-only \impaas\ platform, \market\ \footnote{We do not disclose the real name of the \impaas\ platform to minimize the risk of retaliatory actions from the market operators.}. The platform has operated since late 2016 and grew considerably, in terms of available user profiles, in 2019.
At the time of writing, \market\ provides approximately $260'000$ (and growing) user profiles available for impersonation attacks against Internet users worldwide.
\market\ is a Russian \impaas\ platform reachable from the surface web. This platform is, to the best of our knowledge, %
the first, large \impaas\ operator operating in the underground. On \market, a user profile contains information coming from user systems infected with a credential stealer custom malware acting as a man-in-the-browser. The custom malware enables the exfiltration of cookies, credentials and sniffing of keystrokes, alongside additional environmental and device information that uniquely characterize the user. The \impaas\ platform states user profiles are updated and pushed to the attacker's system in real-time, \review{R1.2, R3.8} 
and that sold user profiles are removed from the listings of profiles available for purchase, although this is difficult to verify empirically, and ethically.\footnote{A proposition is to infect one's own system and purchase back the generated profile to verify its disappearance. As the malware employed by the platform is custom, reproducibility is non-trivial. See also Sec~\ref{subsec:malware_work}.}
An overview of the profile characteristics is provided to browsing customers; profiles with specific characteristics can be searched through the marketplace interface.
From the platform, it is possible to access the list of bought profiles and download the related fingerprint. 
Further, \market\ provides their customers with a custom chromium-based browser plugin and a pre-built version of Chromium for both macOS, Linux and Windows. This bundle can be accessed only after having bought at least one user profile on the platform. %
The plugin comes with the capability of loading fingerprints previously obtained from the acquired profiles and can tunnel the traffic through an attacker-specified SOCKS5 proxy to spoof a victim's geolocation.

\paragraph{Malware customization}
\label{subsec:malware_work}
\review{R1.4, R2.1}
The latest known custom malware employed by \market\ is based on the AZORult malware ~\cite{cylance_2019,gatlan_2020,bisson_2020}.
\market\ reports a recent update (Nov 2019) in AZORult addressing changes introduced in the Chrome browser that appear to have affected the malware functionality. Confirmation of massive phishing campaigns in that period associated with AZORult come independently from Kaspersky and other researchers~\cite{gatlan_2020, bisson_2020, kaspersky_labs_2020}. Note that, start of 2020, AZORult was abandoned by \market\ in favour of a new (and, at the time of writing, still unnamed), custom malware.
Due to the changing nature of the adopted malware, we here only provide a high-level overview of AZORult operations from samples available (at the time of data collection) in the underground and malware repositories. For our analysis we replicated the latest three versions of AZORult (at the time of writing 3.3, 3.4.1 and 3.4.2) in a virtual environment, with the aim of evaluating its overall functionalities and their relevance to \market. %
Malware customization happens through two modules, namely the \textit{builder} and the \textit{C2 server}. The \textit{builder} has the purpose of generating the custom build of AZORult including the URL of the C2 server. 
The \textit{C2 server} module is a ready-to-deploy web service providing an overview of the harvested data and a page for setting up the features of the malware; these features are user-defined, and include the collection of browser history, saved passwords, cryptocurrency client files, Skype history, a customizable regexp-based file grabber targeting user-defined folders on the infected host, and an additional setup for the deployment of a second stage infection on the victim system:
as AZORult removes itself from the system after execution, the second-stage mechanism can allow \market\ operators to obtain persistence on the infected system and further refine the data collection (e.g., to harvest behavioral data over time, see profile updates analysed in Sec~\ref{sec:dataanalysis}). %

\subsection{Platform infiltration}
\label{subsec:infiltration}
Access to \market\ is invite-only, and a valid account is needed to access the listings of available user profiles. Access to the registration procedure is provided through invite codes available to members already active on the platform, provided they spent at least $20$ USD in purchased user profiles. %
To gain access to \market\ we probed several underground forums in which we have a pre-existent foothold, and identified users that claim to be involved with \market. %
As recent evidence suggests that underground platform operators are actively monitoring and blacklisting `rogue' accounts (e.g., performing scraping activities)~\cite{campobasso2019caronte},
we aimed at the collection of several valid accounts prior to data collection to distribute the activity and have `backup' identities to use if some of our accounts were to be blacklisted.
Our search lead us to six members in \hackforum\ and one member in \cardingmarket\ (who claimed to be one of the operators of \market) that were offering free invite codes between December 2019 and March 2020. We contacted them through the private messaging facility of the forums as well as on the messaging board, and obtained valid invitation codes from three of them in \hackforum. From \cardingmarket\ we gained access to an additional eight valid invitation codes using separate (and active) identities on the forum, for a total of eleven \market\ accounts overall. %

\subsection{User profiles on \market}
\label{subsec:userprofiles}

\market\ offers an overview of the available profiles, highlighting the information bundled in that user profile. A view of the interface accessed by attackers is provided in \autoref{fig:listing} and \autoref{fig:advancedsearch} in the Appendix. It is worth to note that, whereas \market\ listings do not readily provide identifying information on the user, the information available on a listing is detailed enough to identify users operating in specific target environments such as a specific organization (e.g., to then perform lateral attacks~\cite{ho2017detecting}). 
\market\ distinguishes between the following information in a user's profile: cookies, resources and fingerprints. 

\smallskip
\noindent\textbf{\cookies.} These are the cookies captured by the custom malware and available for injection toward the respective platforms once the user profile is purchased and enforced by the attacker.

\smallskip
\noindent\textbf{\resources.}
\resources\ are collections of data derived from keylogging activity and probing of browser's local resources, such as the database of stored passwords, and browser history. Some well-known resources (e.g., related to social media platforms, home banking, etc.) are highlighted as \texttt{known resources} by the platform, suggesting that the type of extracted \resources\ is an important information for the attacker to consider. A resource can include multiple data reporting login credentials, answers to security questions, detailed balance info for bank accounts, credit/debit card numbers and holder details. %
\market\ states that the malware extracts \resources\ from infected systems through three main modules: \texttt{FormParser} reads the contents of the form data inputted by the user; \texttt{SavedLogins} gathers credentials saved in the browser's local database; \texttt{InjectScript} implements code injection on the victim's browser on behalf of the attacker, but its operation is unclear and most of the listed profiles do not appear to rely on it. %

 \smallskip
\noindent\textbf{\fingerprints.}
\fingerprints\ provide a collection of the features exposed by a browser when interacting with RBA systems, ranging from technical metadata (user-agents, browser version) to more finely grained features (geolocation, latency, system language, fonts installed, web site device access permissions, etc.)\footnote{Whereas a full list of the probed features is not available from \market\ nor from our analysis (see section ~\ref{subsec:malware_work}), a number of commercial and free solutions could be employed by the \market\ malware to implement reliable fingerprinting of the infected systems. }. \review{R1.4, R3.2} 
Depending on the specific RBA implementation, a service may probe a specific subset of the features characterizing a browser or system. Differently from \resources\ (which are tied to a specific service, e.g. a username/password combination on Amazon), the features collected in an \market\ fingerprint are not bounded to a specific service,  but to the browser environment itself (e.g., available system fonts, or installed plugins). Therefore, these constitute a \textit{pool} of features that can be requested by any service, when available. 
\market\ distinguishes between two types of \fingerprints:
\begin{enumerate}
    \item \textbf{Real fingerprints}: these are directly collected from the victim's device, providing an accurate identity of the impersonated device; albeit rarely available in bots, they appear to be sought after by market users;
    \item \textbf{Synthetic fingerprints}: these fingerprints are generated on the basis of the data collected by the malware. However, accurate `synthetic' fingerprints cannot be generated without user data (e.g, system fonts, plugins installed in a browser, etc.). For this reason, we consider the availability of \resources\ \textit{and} of browser data in a user profile as an indication that the malware is in the position of collecting the necessary data to generate a reliable synthetic fingerprint. 
\end{enumerate}

\subsection{Data collection strategy}
\label{sec:datacoll}

To collect data on \market\ operations, we first consider a number of structural limitations at the core to our sampling strategy:

\begin{enumerate}
    \item[Lim-1] To avoid disclosing our identity to the \market\ operators, we perform the scraping behind TOR. This poses technical limits (as well as ethical concerns) for bandwidth usage.
    \item[Lim-2] We have a limited number of accounts to perform our measurements; aggressive probing risks exposing our accounts to the \market\ operators, and lead to blacklisting.
    \item[Lim-3] Information on \resources\ cannot be accessed in bulk via an API or other requests to \market, but rather have to be requested in limited bundles with separate requests. This explodes the number of requests necessary to obtain \resources\ information on all user profiles on \market.
\end{enumerate}

To address \texttt{Lim-1} and \texttt{Lim-2}, we employ an \textit{ad-hoc} crawler. Initially the crawler was set to work $\approx24$h/day issuing, on average, 15 requests per minute;
despite the relatively low requests volume, this strategy led two of our accounts to be blacklisted, suggesting that \market\ operators may be employing network monitoring solutions to avoid measurement activities. %
Following~\cite{campobasso2019caronte}, we progressively reduced the crawling activity to $\approx 6$h/day. \review{R3.3} 
In the process, an additional three accounts were banned, for a total of five banned accounts. 
It is interesting to note that three of the five blocked accounts were not linked to each other in any way,\footnote{The first two accounts were obtained from a single member of \market\ and active on \hackforum. The other accounts all came through invitation codes generated by either different market members, or released by \market\ operators themselves to different and unrelated accounts we control on \cardingmarket.} suggesting that market operators have kept their crawling-detection efforts high during our activities. To mitigate this problem we employed different strategies to access specific pages and resources to crawl on \market: %
as already noted in~\cite{campobasso2019caronte}, accessing URLs directly (as opposed to via website navigation) may generate anomalies in crawler monitoring systems. For this reason, we operationalised all crawling activities through browser instrumentation, and configured the crawler to mimic activity patterns compatible with those of a human user (e.g. timeouts between requests proportional to the length of the visited webpage, taking breaks, ...). With this final setup, we finally managed to silently crawl the market avoiding the detection and ban of the remaining accounts in our possession. %

While necessary, the above strategy makes it impossible to gather complete information on \resources\ due to the exploding number of requests (\texttt{Lim-3}). %
This results in two datasets: 
\begin{itemize}
    \item \texttt{Full database} includes information on approximately $262'000$ user profiles on \market, including (infection, update) dates, prices, number of browsers for which resources are available, number of collected fingerprints for that user profile, and number of stolen cookies.
    \item \texttt{Sampled database} adds \resources\ information to a random selection of approximately 5\% ($n=13'512$) of the user profiles available on the market.\footnote{This fraction was originally set to 10\%, however approximately half of the selected profiles were removed from the market during the data collection process.}\review{R1.9}
\end{itemize}

The collected data is available for sharing to the research community at \url{https://security1.win.tue.nl}.

\subsubsection{Analysis procedure}

The data analysis in Section~\ref{sec:dataanalysis} is split in two subsections: in Sec.~\ref{sec:dataoverview} we provide an overview of the data collected in the \texttt{Full dataset}, and characterize \market\ operations by looking at its evolution, victim profile characteristics, profile updates, and pricing; in Sec.~\ref{sec:dataresources} we analyse the distribution and effect of \resources\ on pricing, as reported in the \texttt{Sampled database}. Standard sanity checks (e.g. on the regression results presented in Sec~\ref{sec:dataresources}) are performed on all analyses. Reported logarithms are natural logarithms unless otherwise specified.

\smallskip
\emph{Manual resources classification.}
\label{par:manualclass}
To factorize the type of resources reported in \texttt{Sampled database} in the analysis, we provide a classification of each resource in one of six categories. Table~\ref{tab:classdesc}
\begin{table}[t]
\centering
\caption{Categories of resources.}
\label{tab:classdesc}
\begin{tabular}{p{0.12\columnwidth}p{0.56\columnwidth}p{0.2\columnwidth}}
\toprule
Category           & Definition    & Examples \\                             
\midrule
\services        & Platforms providing the delivery of physical (e.g., goods, postage, etc.) or digital (e.g., content streaming, cloud, mail, etc.) services to final users. & Google, PosteItaliane                                    \\
\social          & Platforms to share user generated content. & Twitter, Skype                                                          \\
\money & Platforms enabling direct payments between people or organizations using traditional payment circuits. & CreditUnion, Transfergo\\
\texttt{Cryptoc.}         & Platforms enabling direct payments between people or organizations using cryptocurrency circuits.  &Coinbase, Bittrex     \\
\commerce        & Platforms whose sole purpose is to purchase or book goods/services from one or multiple vendors.  &Amazon, SaldiPrivati            \\
\other           & Platforms that do not match any of the previous categories.                     & Auth0    \\
\bottomrule
\end{tabular}
\end{table}
lists the employed categories and their corresponding definitions. The classification was done manually by one of the authors over $454$ unique platforms for which \resources\ are reported in the dataset. The other author independently classified a random sample of 100 platforms, reaching an agreement score of $89\%$; after review, conflicts were resolved and the classification was updated accordingly. Additional random checks did not reveal any remaining mismatch.

\smallskip
\emph{Ethical considerations and limitations.}
No personally identifiable information is reported in our dataset. IP addresses of victims are masked on the platform, and no detailed information about the victims is available without purchasing a user profile. For obvious ethical concerns, we did not purchase any. Whereas this limits our analysis in that we do not have access to the software bundle provided by \market, and cannot ascertain in detail the quality or operative aspects of the \impaas\ service provided by \market, we are in the position of providing a full evaluation of the data is available to the attacker when browsing for victims.

\section{Data Analysis}
\label{sec:dataanalysis}

Table~\ref{tab:descstats} provides an overview of the collected datasets.
\begin{table}[t]
    \centering
    \caption{Summary statistics of the collected datasets.}
    \label{tab:descstats}
    \begin{tabular}{p{0\textwidth}p{0.002\textwidth}p{0.108\textwidth}p{0.062\textwidth}p{0.062\textwidth}p{0.062\textwidth}p{0.042\textwidth}}
    \toprule
    & & Variable & min & mean & max & sd \\ 
    \midrule
    \multirow{7}{*}{\rotatebox{90}{Full dataset}} & \multirow{7}{*}{\rotatebox{90}{  ($n:262'080$)}} &
  No browsers & 0 & 1.58 & 10 & 1.02 \\ 
  & & No cookies & 0 & 1719.56 & 125198 & 1773.57 \\ 
& & No \textit{real} fprnts & 0 & 0.06 & 17 & 0.32 \\ 
& & Date infection$^\dagger$ & 12-12-17 & 20-11-19 & 16-03-20 & 157.81 \\ 
& & Date updated$^\dagger$ & 01-02-18 & 23-11-19 & 15-09-19 & 156.69 \\ 
& & Country & \textit{char} & \textit{char} & \textit{char} & \textit{char} \\ 
& & Price (USD) & 0.7 & 7.83 & 96 & 7.62 \\ 
\cmidrule(lr){3-7}
\multirow{14}{*}{\rotatebox{90}{Random sample}} & \multirow{14}{*}{\rotatebox{90}{(Dec'17-Mar'20, $n:13'512$)}}&  No browsers & 0 & 1.57 & 8 & 1 \\ 
& & No cookies & 0 & 1782.01 & 26981 & 1735.24 \\ 
& & No \textit{real} fprnts & 0 & 0.12 & 9 & 0.54 \\ 
& & Date infection$^\dagger$ & 28-03-18 & 08-01-20 & 16-03-20 & 40.04 \\ 
& & Date updated$^\dagger$ & 12-11-19 & 14-01-20 & 13-06-20 & 37.62 \\ 
& & Country & \textit{char} & \textit{char} & \textit{char} & \textit{char} \\ 
& & Price (USD) & 0.7 & 8.84 & 63 & 8.17 \\ 
\cmidrule(lr){3-7}
& & No resources & 0 & 31.13 & 1322 & 46.63 \\ 
& & \hspace{0.05in} Crypto & 0 & 0.07 & 18 & 0.6 \\ 
& & \hspace{0.05in} Money Tr. & 0 & 1.66 & 385 & 6.23 \\ 
& & \hspace{0.05in} Social & 0 & 7.95 & 1322 & 18.73 \\ 
& & \hspace{0.05in} Services & 0 & 16.64 & 560 & 24.44 \\ 
& & \hspace{0.05in} Commerce & 0 & 4.66 & 296 & 11.12 \\ 
& & \hspace{0.05in} Other & 0 & 0.15 & 16 & 0.88 \\ 
\bottomrule
\multicolumn{7}{l}{\footnotesize{$^\dagger$: dates are reported in dd-mm-yy format. \textit{sd} in days.}}
\end{tabular}
\end{table}

\texttt{Full dataset}. The data collection spans from Dec 2017 to March 2020, involving approximately $262'000$ user profiles. Most user profiles available on \market\ target only one browser, with the top $5\%$ targeting three browsers. Only 35 user profiles report data for more than six browsers in our data. Cookie distribution is similarly skewed. 
Profiles are distributed globally across 213 countries\footnote{Although there are only 195 recognized countries worldwide, \market\ reports ISO 3166-1 codes, which do not distinguish sovereign nations from dependent territories.}, and prices range from $0.7$ to $96$ USD; 50\% of the profiles cost at most 5 USD, whereas the priciest 5\% are priced above 20 USD. %

\texttt{Sampled dataset}. This dataset reports data on $5.2\%$ of the profiles available on \market\, spanning from March 2018 till March 2020 ($n=13'512$). For this dataset, we collected detailed information regarding the available \texttt{resources}. As this is a random sample, values are distributed similarly to \texttt{Full dataset}. Additionally, we extract information on the number and type of resources available for each profile. The average profile has upwards of 30 resources; most resources are of type \services, whereas \social\ and \commerce\ are less common. \crypto\ and \money\ resources appear to be the least numerous in a profile.

\subsection{Overview of \market\ operations}
\label{sec:dataoverview}
To provide an overview of the \impaas\ operations conducted in the market we first look at the full dataset summarized at the top of Table~\ref{tab:descstats}.
Interestingly, we find that approximately $12\%$ of all profiles are \textit{not} associated to a browser on the victim's system.\footnote{Note that all profiles without browser data also do not, by definition, report any data on cookies or fingerprints.} As these profiles do not allow for impersonation attacks under the \impaas\ model, we remove those from further analysis. %
Relative to the number of user profiles, the number of available \textit{real} fingerprints is surprisingly low, with only $4.3\%$ of the available profiles having at least one. Note however that this refers only to \texttt{Real fingerprints} collected by the malware, not the \texttt{Synthetic fingerprints} that can be synthesized from user data (ref. Sec.\ref{subsec:userprofiles}). Nonetheless, this suggests that (real) fingerprints, available browsers, cookies, and resources could be the driving force behind \market\ activities.

\begin{figure*}[t]
\centering
\includegraphics[width=0.49\textwidth]{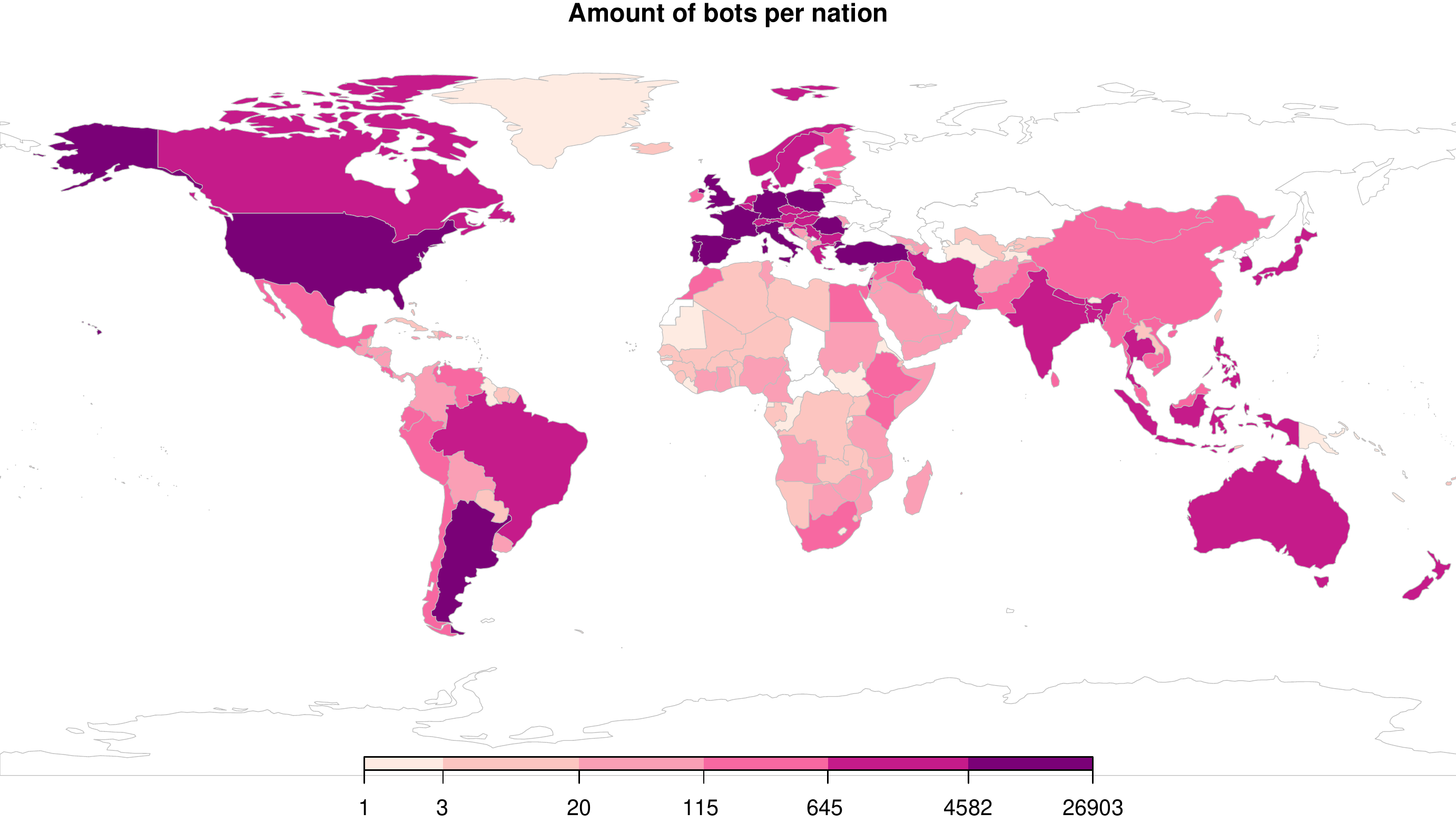}
\includegraphics[width=0.49\textwidth]{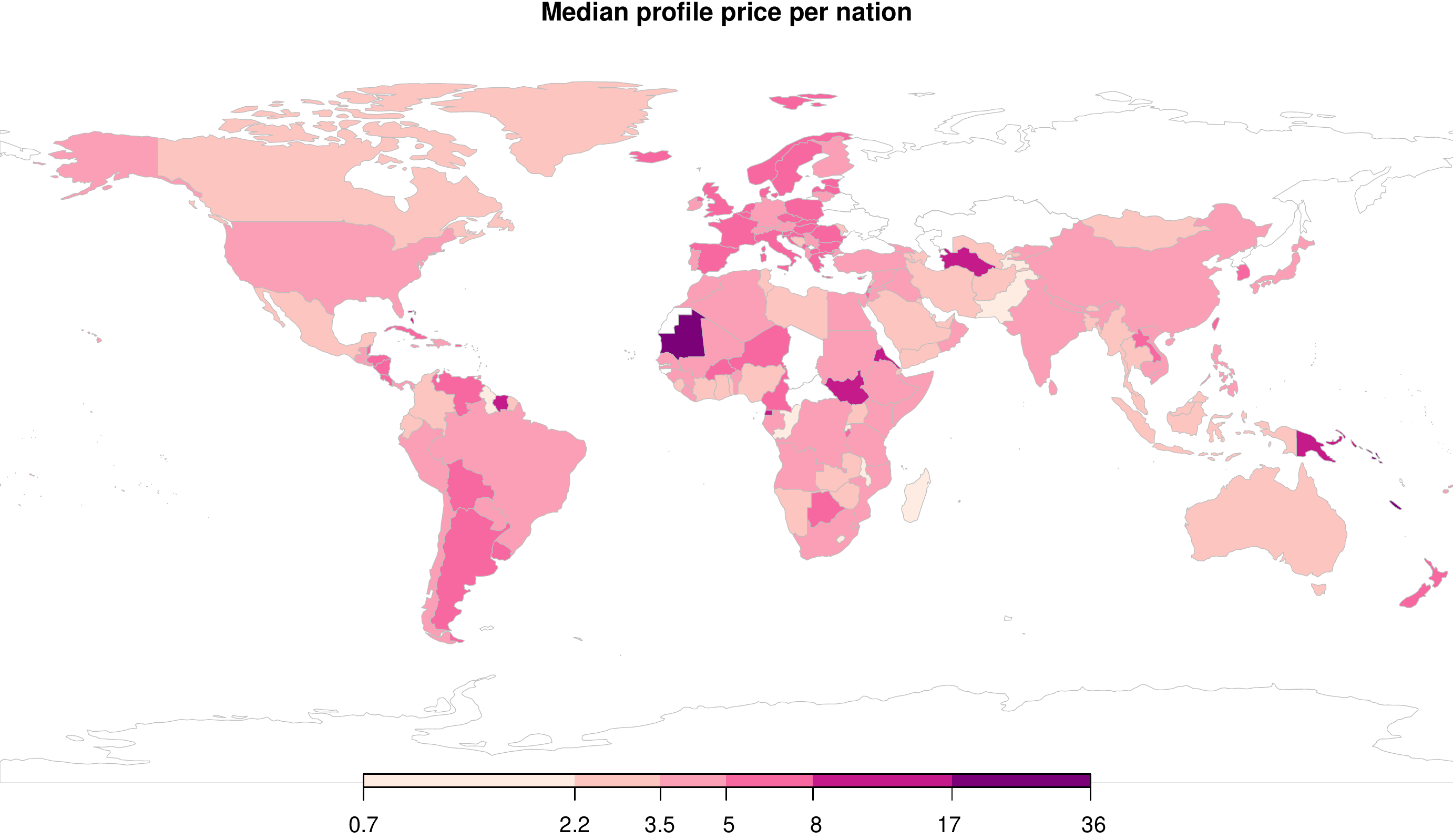}
\caption{Global distribution of user profiles (left) and their median price (right) on \market}
\label{fig:geo}
\end{figure*}

Figure~\ref{fig:geo}
provides an overview of the geographic distribution of the user profiles available on \market\ and their median price per country. Most of the profiles belong to users in the United States of America and Europe, with a high fraction of EU countries showing volumes similar to those of the US. Users in Asian and African countries are comparatively less affected. As commonly seen in Russian cybercrime markets~\cite{allodi2017economic}, \market\ does not provide profiles for users in Russia, Ukraine, Belarus, and Kazakhstan (CIS countries). Furthermore, with the exception of Chad and the Central African Republic, the CIS countries appear to be some of the only unaffected countries, globally. Overall, median prices appear to vary from country to country rather than at a macro-regional level. For example, EU median prices seem to be higher in Spain ($m=9.55, sd=9.07$) and GB ($8.3, sd=7.5$) than in Germany ($m=7.21, sd=8.21$) or Finland ($m=6.96, sd=6.68$). A set of Wilcoxon Rank-Sum tests evaluating the alternative hypothesis that SP and GB profile prices are higher than DE and FI ones confirms this observation ($p<0.0001$). The high median price in Mauritania (26 USD) is caused by only one profile (with no fingerprints, two browsers, and four thousand cookies) available for that country.

\begin{figure}[t]
\centering
\includegraphics[width=0.9\columnwidth]{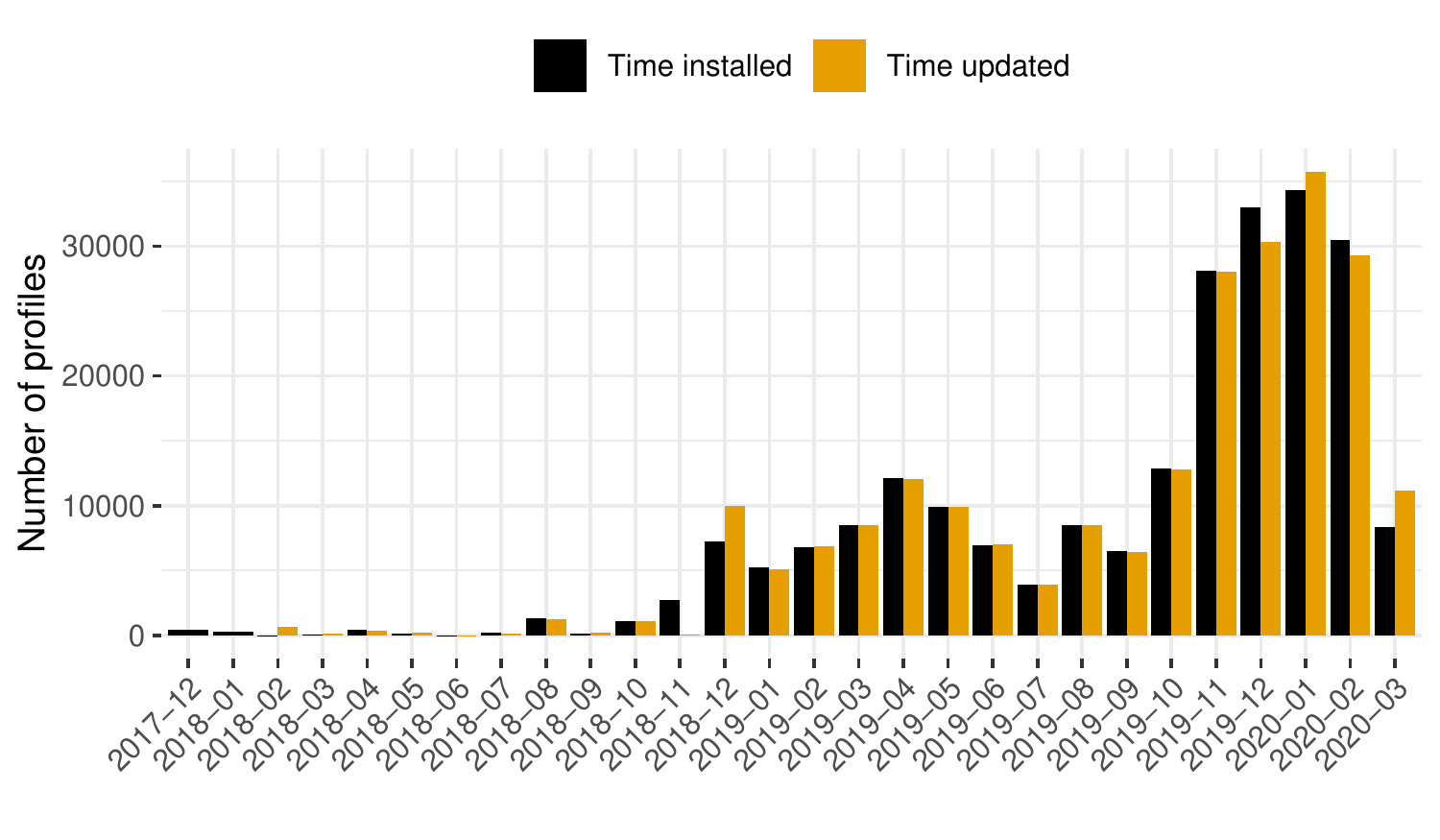}
\caption{Progression of available user profiles over time.}
\label{fig:botsovertime}
\end{figure}

\begin{figure}[t]
    \centering
    \includegraphics[width=0.9\columnwidth]{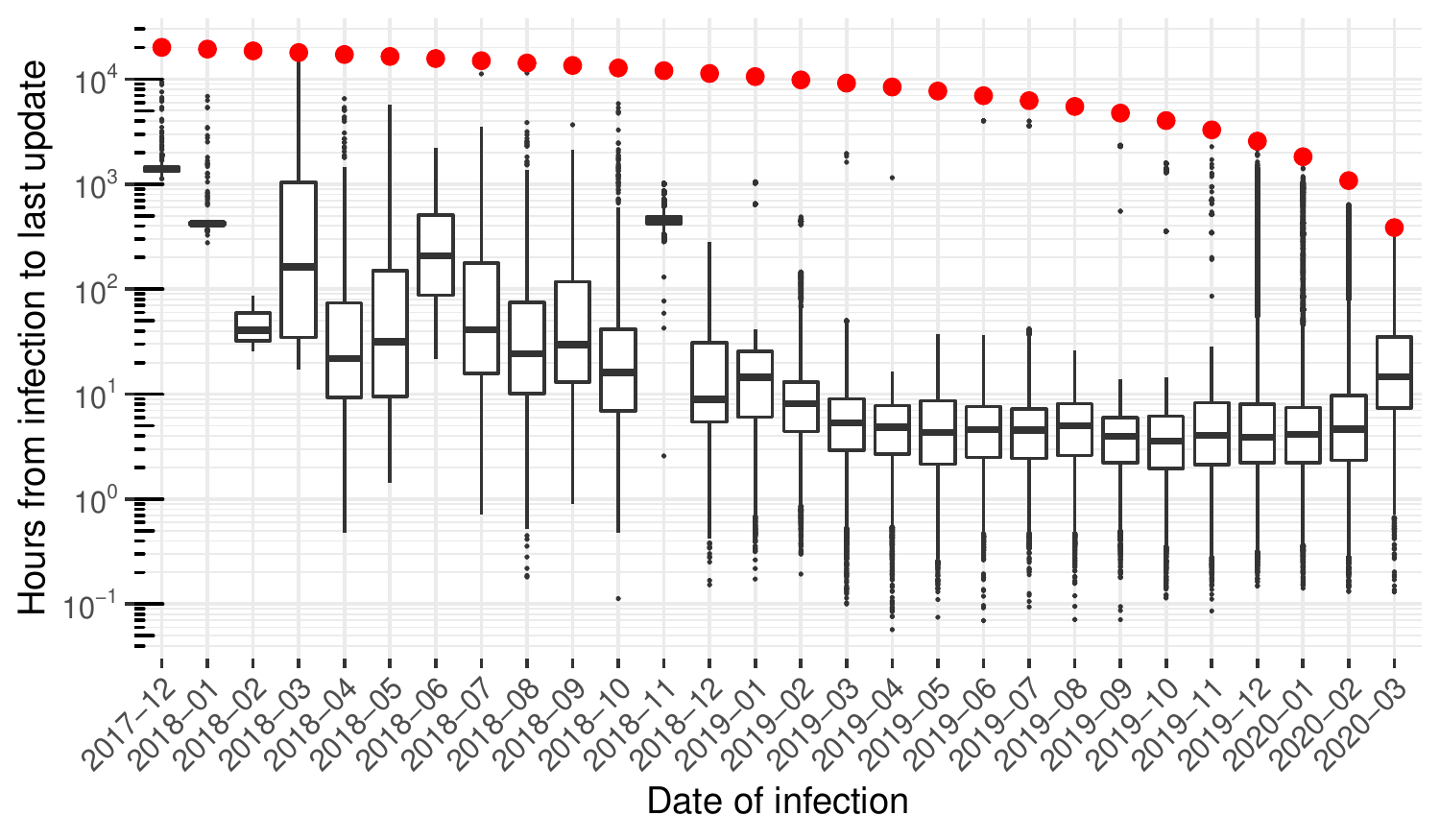}
    \caption{Time between infection and last profile update (in log scale).}
    \label{fig:diffs}
\end{figure}

The rate of appearance of new and updated user profiles on \market\ is depicted in Figure~\ref{fig:botsovertime}.
A clear upwards trends in terms of number of available profiles is visible, with a large jump in available profiles in November 2019 (coinciding with the 2019 spike in phishing campaigns distributing AZORult~\cite{gatlan_2020,bisson_2020,kaspersky_labs_2020}).
\review{Fig.~\ref{fig:pricesboxplot} shows the boxplot distribution as requested by R3.5. Fig.~\ref{fig:pricesboxplot} and \ref{fig:prices} seem largely equivalent and we'd prefer to keep the moving average.} Overall, in Figure~\ref{fig:botsovertime} we observe a sustained rate of new (black bar) and updated (orange bar) profiles, suggesting that the platform is systematically updating existing profiles, while adding new ones to the platform portfolio. We further investigate the time passing between time of infection and (last) profile updates; Figure~\ref{fig:diffs}
shows the boxplot distribution of time passed between the infection and the last update received by the platform, plotted against the date of installation; in red, the upper bound of the maximum possible time in between.\review{R1.7} Overall the distribution appears relatively stable, with a median update time ranging between ten hours and four days. Unsurprisingly, recently acquired profiles are updated only after a few hours from acquisition; overall, the distribution suggests that profiles are kept updated on average over an extended period of time, ranging from a few days, to several months at the extreme of the distribution.

\subsubsection{Analysis of profile values}

\begin{figure}[t]
\includegraphics[width=0.9\columnwidth]{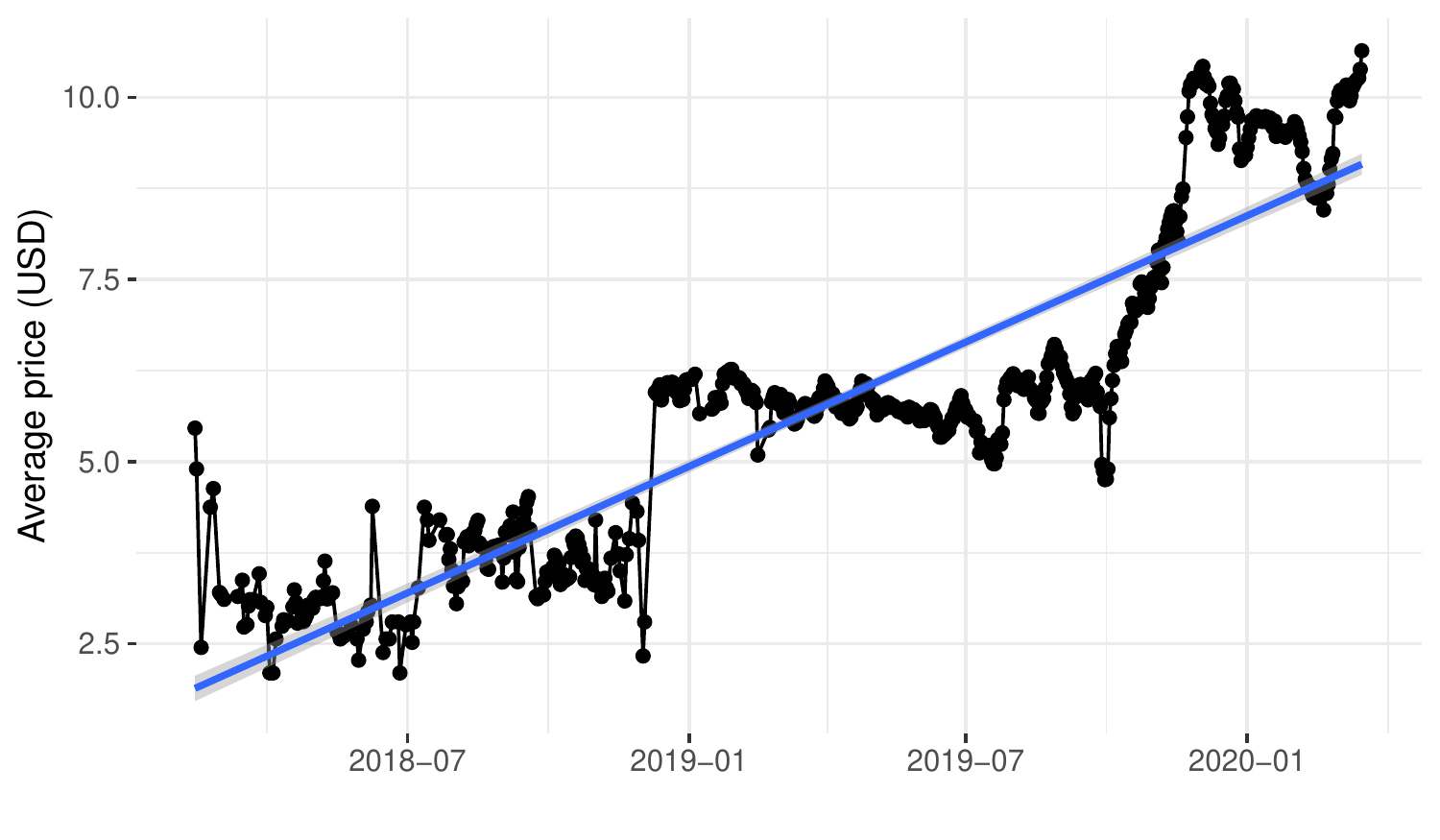}
\caption{Weekly moving average of user profile prices.}
\label{fig:prices}
\end{figure}

\begin{figure}[t]
    \centering
    \includegraphics[width=0.9\columnwidth]{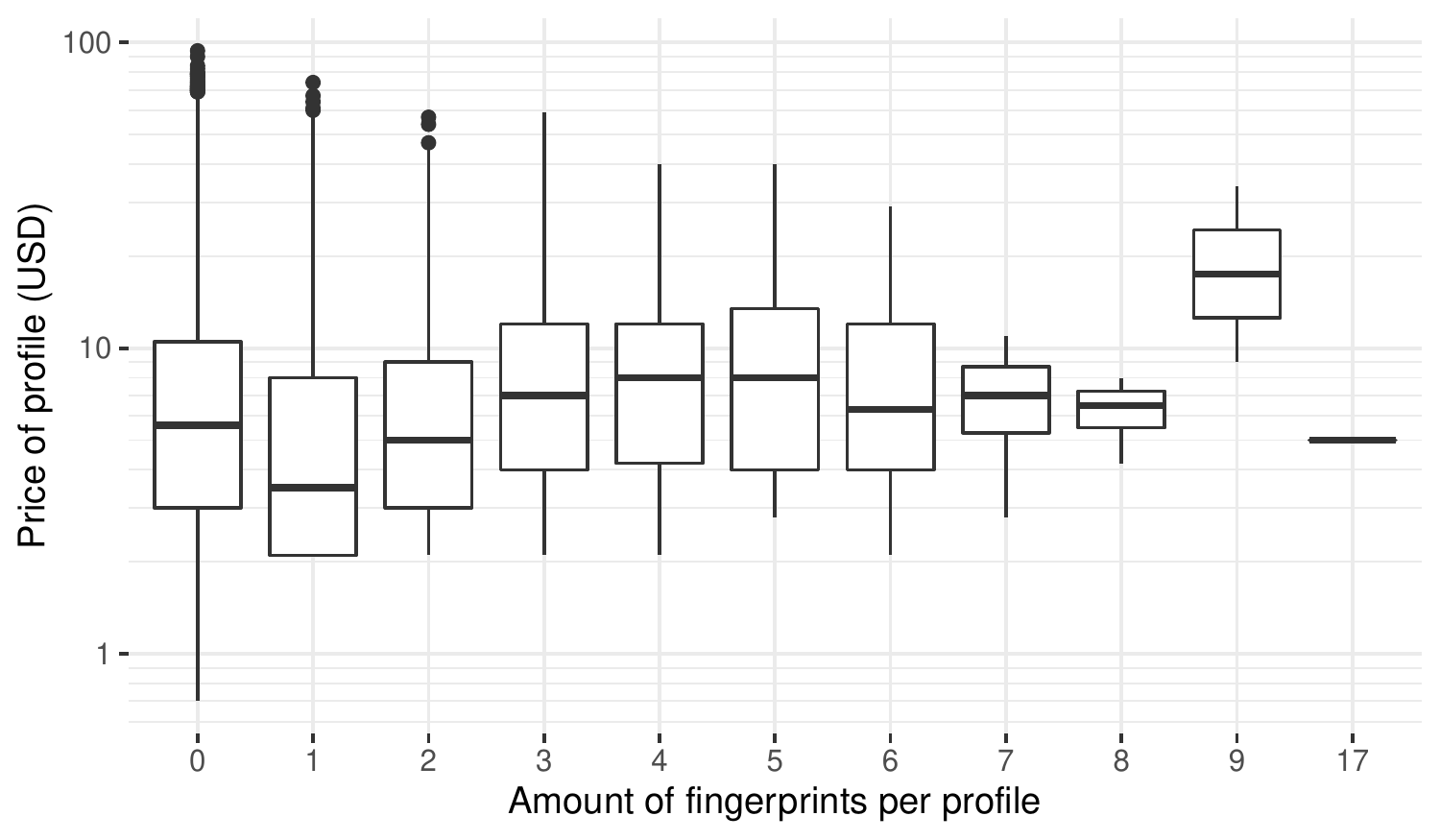}
    \caption{Relation between the amount of fingerprints available and the average price of user profile (in log scale).}
    \label{fig:nofingerprints}
\end{figure}

Figure~\ref{fig:prices}
reports the moving average of user profile prices as a function of time. 
The value of the traded profiles steadily increases as time passes, a signal of growth of the platform. In particular, profile prices seem to have doubled since November 2019, perhaps as an effect of the updated malware released in that period discussed at the start of this Section. %
\autoref{fig:nofingerprints}
reports the relation between the number of available \fingerprints\ in a profile and its price. The effect of an increased number of available fingerprints is, albeit positive, very limited. The average price seems to stabilize around the median value of 5 USD regardless of the number of fingerprints available in the profile, suggesting once again that other variables could be at play. We find no correlation between number of available browser and number of cookies and prices. This is not surprising, as these dimensions express little in terms of \textit{which} identities of the victim the attacker may affect.

\begin{figure}[t]
    \centering
    \includegraphics[width=0.9\columnwidth]{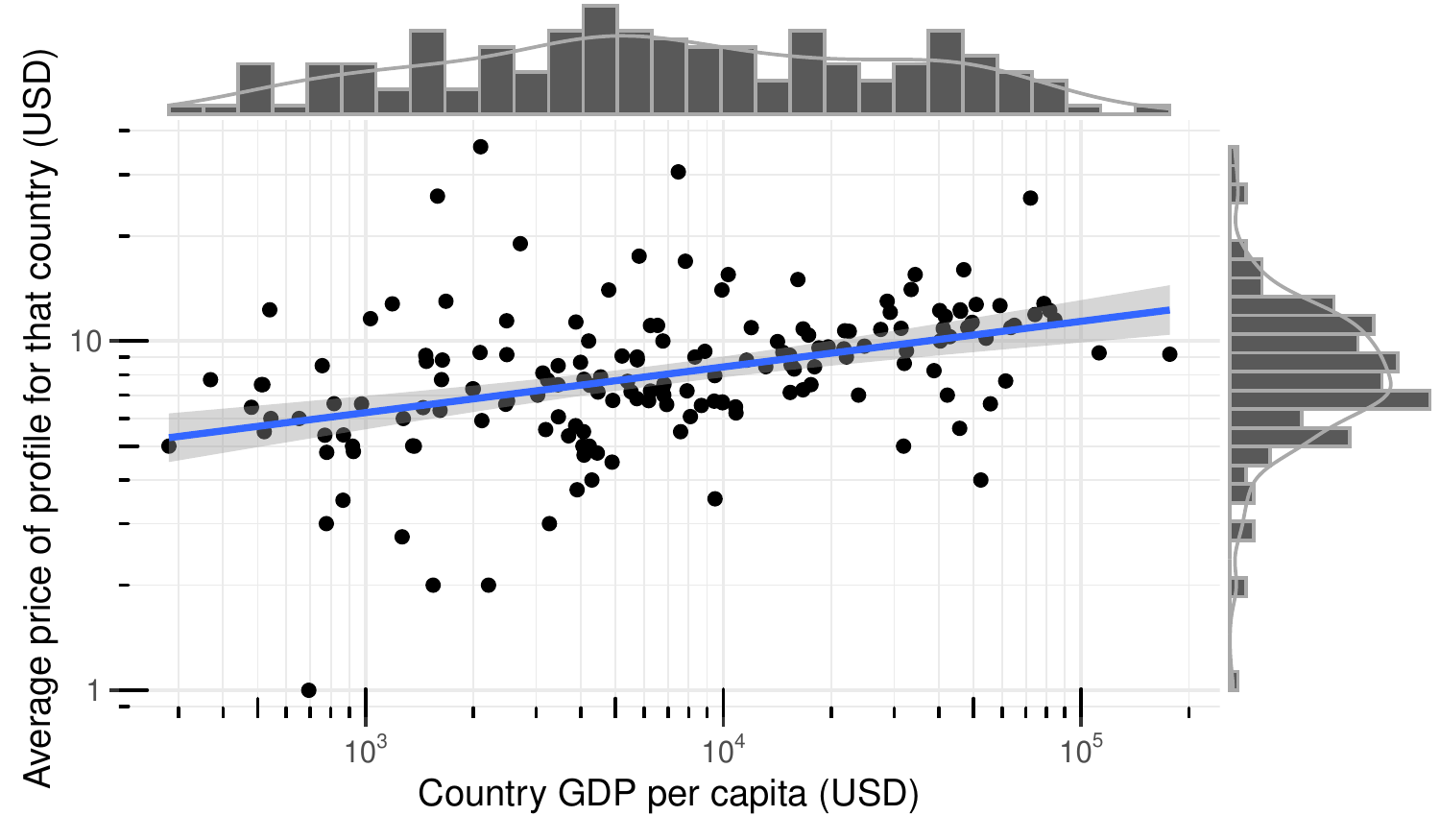}
    \caption{Relation between GDP per capita and average price of user profiles in that country (in log scale).}
    \label{fig:wdicountry}
\end{figure}

To further look at factors that may determine the value of a profile, we look at the impact of the geographic location to which the profile is linked. To do so, we investigate the relation between (log-transformed) profile prices and the wealth of the country in which the profile is located, expressed in terms of (log) GDP per capita (as reported by the World Development Indicators~\cite{wdi}). The intuition is that, the more `valuable' a target is perceived to be, the greater the value of the corresponding profiles might be. \autoref{fig:wdicountry}
reports the analysis.
A positive and statistically significant correlation emerges, suggesting that profile prices are indeed correlated to the wealth of the respective country, perhaps a sign of the perceived value of that user profile ($corr=0.4, p<0.001$)\review{R3.7, R4.2}.

We note that some user profiles on \market\ appear to be discounted at a rate of 30\%. We do not find a clear-cut effect explaining which profiles are likely to be discounted (Table~\ref{table:coeff_sale} in the Appendix).\review{R1.3}

\subsection{The impact of \resources\ on profile pricing}
\label{sec:dataresources}

We first look at the distribution of resource types in the \texttt{Sampled dataset}. As for the \texttt{Full dataset}, we remove from further analysis profiles that aren't associated to at least a browser of the victim's system and, in addition, profiles that don't contain any stolen resource, limiting the size of the dataset to $n=12'052$. %
Table~\ref{tab:categoriescount}
\begin{table}[t]
    \centering
    \caption{Type of resources per user profile.}
    \label{tab:categoriescount}
    \begin{tabular}{lr}
    \toprule
         Resource type& no. profiles ($n=12'052$)\\
        \midrule
            Cryptocurrency&236\\
            Money Transfer&3109\\
            Commerce&5'066\\
            Social&8'111\\
            Services&11'167\\
            Other&548\\
\bottomrule
\end{tabular}
\end{table}
provides an overview of the distribution of user profiles per category. Note that a profile can have resources that belong to more than one category. Overall,  \services\ is the most commonly available resource type across user profiles. Resources in the \social\ and \commerce\ categories are also common, with respectively about 70\% and 40\% of user profiles with resources in these categories. Approximately 25\% of the profiles have data for banking and payment accounts; by contrast, less than 2\% of user profiles have resources in the \crypto\ category. Only $4.5\%$ of the resources in our dataset were classified as \other, indicating that the proposed classification covers the vast majority of resource types in \market.

\begin{figure}[t]
    \centering
    \includegraphics[width=0.9\columnwidth]{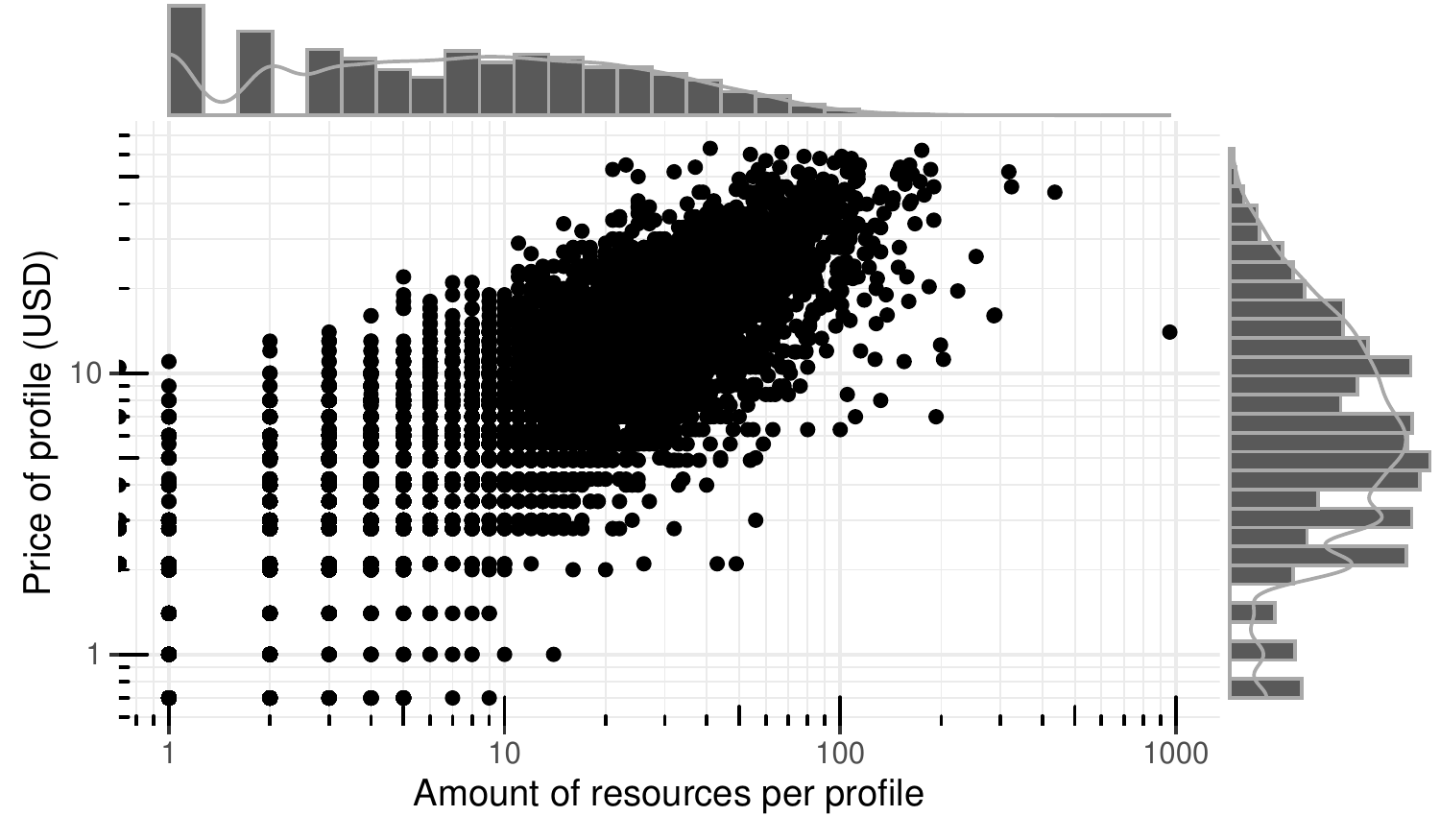}
    \caption{Relation between the amount of resources available and the average price of user profile (in log scale).}
    \label{fig:noresources}
\end{figure}

\begin{figure}[t]
    \centering
    \includegraphics[width=0.9\columnwidth]{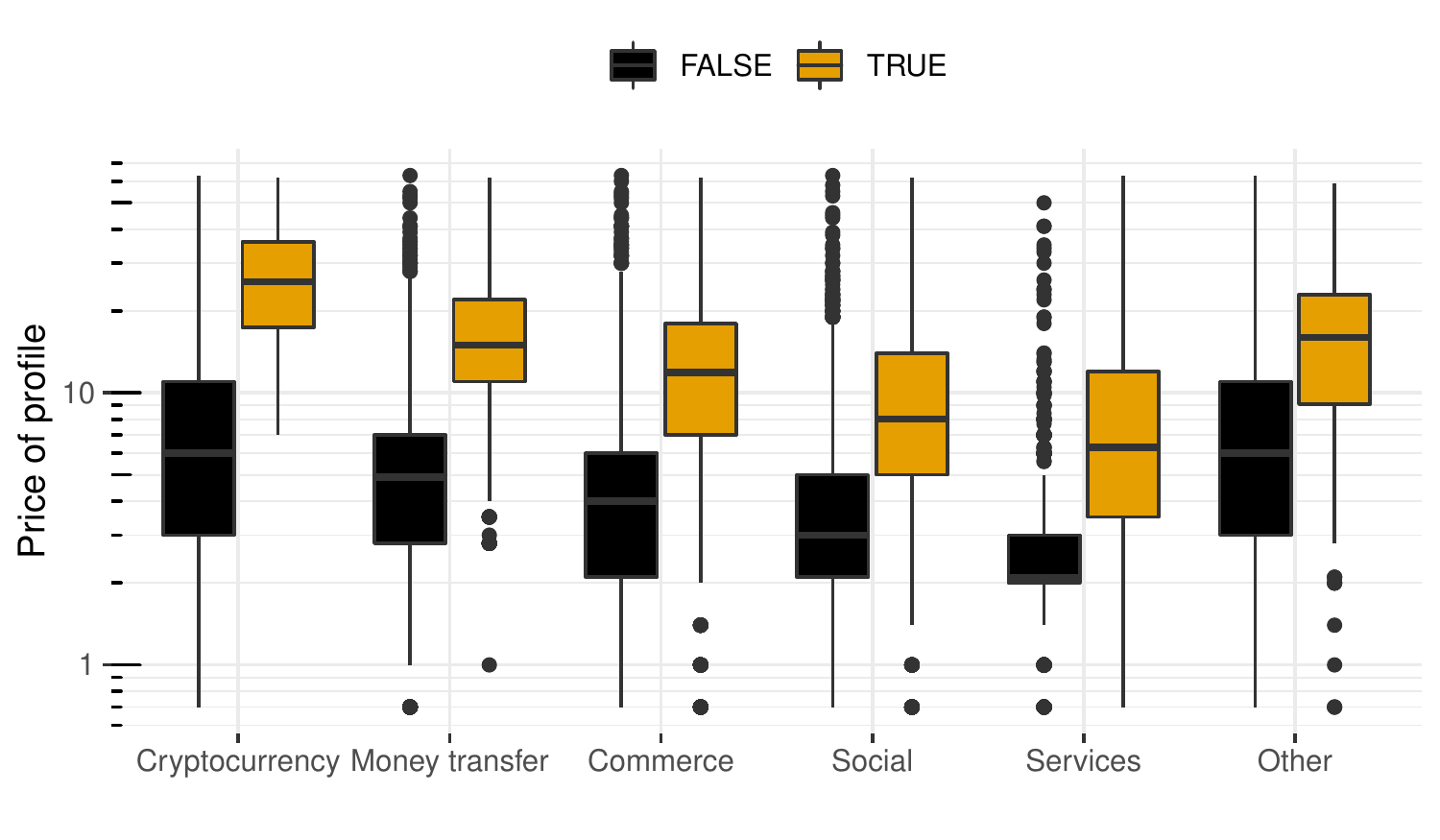}
        \caption{Profile price variation according to the presence of not of resources of a certain category (in log scale).}
    \label{fig:pricepercategory}
\end{figure}

\autoref{fig:noresources} 
provides a first overview of the relation between the number of resources available in a profile and the associated price.  A clear correlation emerges. The depicted linear log-log relation indicates negative marginal returns for each added resource, meaning that every additional resource added to a profile provide an increasingly smaller, albeit positive, added value to the profile. Further exploring the impact of resources on pricing, Figure~\ref{fig:pricepercategory}
shows the impact of the presence of resources in any specific category on the value of a user profile. Note that, because each profile can contain resources of more than one category, one cannot isolate the relative importance of each category here. However, the comparison shows how, on average, a profile that contains (also) resources in that category is priced versus other profiles that do not have it. This is meaningful as the categories show relatively low correlations (reported in Table~\ref{tab:corr} in the appendix). On average, profiles that include \crypto\ resources seem to be the most valuable. \money\ and \commerce\ resources belong to profiles of approximately the same value, whereas profiles with \social\ and \services\ are the least valued in \market. By comparing the relative `jump' introduced by the addition of each category, one can further evaluate the added value, on average, of having a resource of that type. In this respect, \other\ appears to be the least `impactful' category, as the appearance of a resource of this type is related to the smallest relative increase in price, on average, in a profile. On the contrary, \crypto\ and \money\ resources cause the highest jump in profile value, passing from a median value of approximately 7 USD to more than 20. Other categories show less extreme changes in price. Overall, we find that resources associated to financial platforms and services appear to have the highest impact on the value of a profile, with \social\ and \services\ being the least valued. On the other hand, the addition of resources of any category appears to have a positive impact on the value of a profile.

To formally evaluate this relation, we build a set of linear regression models to quantify the effect of different profile features on profile values in \market. To evaluate the effect of each factor independently, and monitor its relation to other characteristics of a user profile, we define the following nested models with response variable $y=price$ (the error term $\epsilon_i$ is omitted for brevity):
\begin{align}
\nonumber M1: y_i = \beta_0 &+ \beta_1 \mathtt{Real\ fingerprints_i} \\
\nonumber M2: y_i = \ldots  &+ \beta_2 \log(\mathtt{GDP_i}) \\
\nonumber M3: y_i = \ldots  &+ \beta_3 \crypto_i + \beta_4 \money_i \\
\nonumber &+  \beta_5 \commerce_i + \beta_6 \social_i\\ 
\nonumber        &+  \beta_7 \services_i+ \beta_8 \other_i \\
\nonumber M4: y_i = \ldots &+ \beta_9 \mathtt{resources_i}
\end{align}
where $\beta_0$ is the intercept, $\mathtt{Real\ fingerprints}$ is the number of fingerprints embedded in that user profile, $\log(\mathtt{GDP})$ is the natural logarithm of the gross domestic product per capita for the country associated with the user profile, $\{\crypto \ldots \other\}$ are dummy variables representing the presence of resources of the corresponding category, and $\mathtt{resources}$ is the overall number of resources in that profile (irrespective of category). 

Regression results are summarized in \autoref{table:coefficients}. To evaluate the effects of profile characteristics on full prices we remove profiles `on sale' from the dataset. Table~\ref{table:coefficientsfull} and Table~\ref{table:coefficientsfull_wsale} in the appendix report, respectively, a full breakdown of the variables' impact on the prediction, and the regression results for all data points including profiles on `sale'; both tables report results quantitatively and qualitatively in line with those reported in \autoref{table:coefficients}.
\begin{table}[t]
\begin{center}
\caption{Regression analysis on prices of user profiles.}
\label{table:coefficients}
\begin{tabular}{l c c c c }
\toprule
 & Model 1 & Model 2 & Model 3 & Model 4 \\
\midrule

$\beta_0$                                & $10.41^{***}$ & $-12.11^{***}$ & $-5.57^{***}$ & $-3.70^{***}$ \\
                                         & $(0.11)$     & $(1.21)$     & $(0.81)$     & $(0.63)$     \\
\texttt{\textit{Real} Fngrpr}            & $0.55^{***}$ & $0.69^{***}$ & $1.31^{***}$ & $1.11^{***}$ \\
                                         & $(0.16)$     & $(0.16)$     & $(0.10)$     & $(0.07)$     \\
$\log{(GDP)}$                                 &              & $2.29^{***}$ & $0.46^{***}$ & $0.42^{***}$ \\
                                         &              & $(0.12)$     & $(0.08)$     & $(0.06)$     \\
\texttt{Crypto}                          &              &              & $13.62^{***}$& $10.12^{***}$\\
                                         &              &              & $(0.44)$     & $(0.34)$     \\
\texttt{Money Transfer}                  &              &              & $8.86^{***}$ & $6.20^{***}$ \\
                                         &              &              & $(0.16)$     & $(0.13)$     \\
\texttt{Commerce}                        &              &              & $5.06^{***}$ & $3.22^{***}$ \\
                                         &              &              & $(0.15)$     & $(0.12)$     \\
\texttt{Social}                          &              &              & $3.44^{***}$ & $1.68^{***}$ \\
                                         &              &              & $(0.15)$     & $(0.12)$     \\
\texttt{Services}                        &              &              & $3.95^{***}$ & $2.31^{***}$ \\
                                         &              &              & $(0.29)$     & $(0.22)$     \\
\texttt{Other}                           &              &              & $4.22^{***}$ & $0.89^{***}$ \\
                                         &              &              & $(0.31)$     & $(0.24)$     \\
\texttt{Resources}                       &              &              &              & $0.10^{***}$ \\
                                         &              &              &              & $(0.00)$     \\

\midrule
R$^2$             & <0.01         & 0.05         & 0.65         & 0.79  \\
Adj. R$^2$        & <0.01         & 0.05         & 0.65         & 0.79  \\
Num. obs.         & 7123          & 7123         & 7123         & 7123  \\
\bottomrule
\multicolumn{5}{l}{\scriptsize{$^{***}p<0.001$, $^{**}p<0.01$, $^*p<0.05$}}
\end{tabular}
\end{center}
\end{table}
Overall, the coefficient estimates appear stable across the models, with the exception of $\beta_2$ ($\log(\mathtt{GDP})$), that becomes less important on the estimation of the dependent variable $\mathtt{price}$ as the types of \resources\ are added to the model. The change ranges from an expected increase of $0.2$ USD in profile value for every $10\%$ increase in GDP ($\beta_2=2.29$ in M2, $2.29\times\log(1.10)=0.22$), to a relatively smaller ($0.04$ USD) price increase when all resource categories are added in the model. 
This indicates that some resource categories may appear more frequently for high GDP countries than for others; with reference to Table~\ref{table:coefficientsfull} in the appendix, it appears that resources of type \money\ and \commerce\ tend to appear more often in wealthy countries, as most of the effect of the GDP variable disappears when this category is accounted for, while the opposite effect emerges when \social\ resources are included in the model. Additional resource categories have modest effects on the GDP coefficient estimate. As resource categories are added to the model, the impact of the number of fingerprints increases, passing from a $0.55$ USD increase in expected profile value for each additional fingerprint in the profile ($\beta_1=0.55$) to a $1.31$ USD increase estimated by M3. This suggests a positive joint effect of the number of fingerprints in a profile, and the number of platforms with resources an attacker can employ to impersonate a victim. All resources have a positive effect on the value of a user profile with \crypto\ and \money\ having the highest impact, increasing the expected value of $13.62$ and $8.86$ USD respectively when available. Following this trend, \commerce\ shows a relatively large effect as well, increasing the profiles' expected value of $5.06$ USD. These findings may not come as a surprise, and may indicate that \market\ customers may be primarily aiming at economical profit (supporting insights from observing \market\ customers discussing on a dedicated Telegram channel, see Sec.~\ref{subsec:examples} for an informal report). Finally, the effect of the number of resources in M4 is significant and positive; interestingly, its addition decreases the effect of the single resource categories, confirming the intuition that the more platforms an attacker can impersonate, the higher the value of the profile.\footnote{Driven by observations in Chen et al.~\cite{chen2009large}, who identified cookies as having a key role in behavioral fingerprinting practices, we find that in terms of profile pricing the availability of cookies does not show a statistically significant effect (Anova $F1.92,\ p=0.17$) in our dataset, suggesting that cookies do not play a central role in impersonation attacks as driven by \market.}

In all, resource types appear to explain the majority of the variance in the model, with \money\ accounting for a jump in more than 30\% in the model (adjusted) $R^2$ when compared with previous model. The complete model explains most of the price variance in our dataset ($R^2=0.79$), suggesting that the model provides an appropriate description of the features determining user profile values in \market.

\section{Discussion}
\label{sec:discussion}

In this paper we presented the \impaas\ model as a novel threat enabling attackers to perform user impersonation at scale. \impaas\ is supported by an emergent
criminal infrastructure that controls the supply chain of user profiles, from system infection to profile acquisition and commodification. Whereas traditional impersonation attacks relying solely on stolen credentials are greatly mitigated by risk-based and two-factor authentication systems, the capability of seamlessly reproducing a user's `appearance' to an authentication system allows attackers to systematically compromise accounts of multiple users, across multiple platforms.

Whereas Thomas et al. already suggested that user profiling could be used to bypass modern authentication systems~\cite{thomas2017data}, in this work we provide evidence of an emergent \textit{as-a-service} impersonation model that appears to be rapidly expanding. The profile value analysis provided in Section~\ref{sec:dataresources} suggests a mature pricing model, which may indicate that the analyzed platform operations are of stable, predictable quality, and likely to expand in number. Overall, the analysis of the available user profiles on \market\ and the reportedly widespread adoption of \textit{info-stealer} malware such as AZORult in phishing campaigns~\cite{gatlan_2020,krebs_2020} provide further supporting evidence of the growth of this threat model.
Our analysis of \market\ allows us to further quantify the relative effects of different resources on the value of a user profile. Interestingly, albeit perhaps not surprisingly, we find that profile values show a significant correlation with the wealth of the country (expressed in terms of GDP) associated to that profile; this suggests that attackers looking to impersonate and, likely, monetize user profiles assign a greater value to profiles likely to give access to greater financial resources (e.g., bank balances or valid credit cards). Interestingly, this effect is significantly reduced by the presence of \commerce\ resources in a profile, perhaps due to the prevalence of e-commerce platforms in wealthy countries. Nonetheless, other resource categories have a clear impact on the overall valuation of a user profile, with \crypto\ and \money\ resources driving most of the value. \textit{Real} fingerprints (those derived directly from the device, rather than being synthesized by the \impaas\ platform using the profile's metadata) available in a profile also add value to the user profile. Our analysis suggests that each \textit{real} fingerprints adds about $0.55$ USD to the value of a user profile, and up to $1.31$ USD when considered jointly with the available resources, suggesting that the \textit{modus operandi} enabled by \impaas\ described in Figure~\ref{fig:fingerprinting} is supported by the platform operations.

Importantly, our analysis allows us to put a number on the value of user information to attackers, contributing to the literature on the subject. A user's `virtual identity' seems to be worth between less than 1 USD and approximately 100 USD. This value changes significantly depending on the wealth of the country where the user (appears to be) located; a rule-of-thumb indication seems to be that for a tenfold increase of the a user's `expected wealth' (approximated by a country's GDP), a profile value increases on average by approximately $1$ USD. 
Cybercriminals seem to particularly value profiles with access to \crypto\ and \money\ platforms, whose prices are respectively 10 and 6 USD higher than profiles with no access to platforms of these types. To put this in perspective, these represent respectively a 150\% and 90\% markup over the price of the average profile, a clear sign of the relevance of resources of these types for cybercriminal activities. By contrast, access to \social\ and other services does not seem to be (in comparison) highly valued by cybercriminals.

\subsection{Implications for victimization} 

The systematization of impersonation attacks enabled by the \impaas\ model allows attackers to select and target specific victim profiles, and to automate the attack procedure by means of dedicated software bundles replicating a victim's browsing conditions on the environment of the attacker. Differently from traditional phishing-based attacks, \impaas\ provides an attacker with access to several platforms on which a user is active, effectively allowing the attacker to both mitigate security measures (e.g., by monitoring email for authentication codes or activity notifications), as well as extending the attack surface to different services (e.g. banking, social, etc.).

Attackers leveraging an \impaas\ platform can rely on an automated source for credentials to conduct sophisticated attacks at scale. In addition to obtaining access to banking websites, cryptocurrency exchange platforms and e-commerce websites, an attacker may  compromise multiple accounts to gain control over the identity of the victim. 
The capability of selecting victim characteristics before the acquisition of a profile is also a potential enabler of targeted attacks against organizations or communities for which a victim is an employee, or a registered member. The attacker may employ that advantage point to facilitate lateral movement attacks, for example targeting colleagues or family members of the victim by using their legitimate contact details. Furthermore, the attacker could integrate additional information about a victim gathered through the accessed platforms (part of a corporation, subscription to meeting websites, etc.) to further escalate the attack to other victims. %

\subsection{Examples of (alleged) criminal operations enabled by \market}
\label{subsec:examples}
To informally investigate how attackers are weaponizing capabilities enabled by \impaas, we collected a number of examples provided by users of \market\ on a Telegram channel linked to the platform, to which we have gained access through \market. 
Many attacks reported there appear to focus on \money\ and \commerce\ services. For example, a user shared that they were (allegedly) able to cash-out from a US bank using a synthetic fingerprint acquired on \market\, and with the support of a geographically accurate SOCKS5 proxy. The user further suggested to rely on \texttt{911.re} as the marketplace where to buy proxies linked to specific ZIP codes and/or ISPs.
On a similar line, a second user reported to have managed issuing a new debit card on behalf of the victim, with the aim of cashing it out through ATMs. 
Interestingly, some \market\ users report performing multi-stage attacks deployed through the obtained user profiles and exploiting multiple platforms. For example, an attacker describes setting filters to a victim's email mailboxes accessed through the victim's user profile, with the aim of hiding notifications from Amazon related to purchases the attacker made using the victim's Amazon account.

Overall, whereas of course none of these examples can be verified and the threats described above are not new per-se, the mix of infrastructural support for profile acquisition, selection, and enforcement enabled by \impaas\ opens to the \textit{systematization} of threat scenarios such as the ones described above, on a global scale.

\section{Conclusion}
\label{sec:conclusion}

In this paper we presented the emergence of the \emph{Impersonation-as-a-Service}
criminal infrastructure, which provides user impersonation capabilities for attackers at large. \impaas\ allows attackers to bypass risk-based authentication systems by automatically simulating the victim's environment on the attacker's system. In this study we characterise the largest currently operating \impaas\ infrastructure, \market, by performing an extensive data collection spanning more than 260 thousand stolen user profiles collected worldwide. \market\ infiltration and data collection required substantial efforts to collect multiple accounts, needed to fine-tune the data collection as platform operators seemed to monitor crawling activities and blacklist related accounts. From our analysis, \market\ emerges as a mature, expanding infrastructure with a clear pricing structure, suggesting a well-established criminal business model. \emph{Impersonation-as-a-Service} represents an additional component of the cybercrime economy, providing a systematic model to monetize stolen user credentials and profiles.

\smallskip
\textbf{Lesson learned.} Our data collection efforts provide supporting evidence that underground platform operators are actively monitoring crawling activities, and take measures to limit them. This may prevent future research activities and significantly impact the possibility of designing large-scale studies studying cybercriminal online venues. Specific sampling strategies and analysis techniques will have to be devised to further develop research in this domain.

\section*{Acknowledgements}
This work is supported by the ITEA3 programme through the DEFRAUDIfy project funded by Rijksdienst
voor Ondernemend Nederland (grant no. ITEA191010).

\bibliographystyle{acm}
\balance
\bibliography{biblio,bib,short-names,security-common}

\newpage

\appendix
\balance
\section*{Appendix}

\section{Market features}
\begin{figure}[hbp!]
\includegraphics[width=0.8\columnwidth]{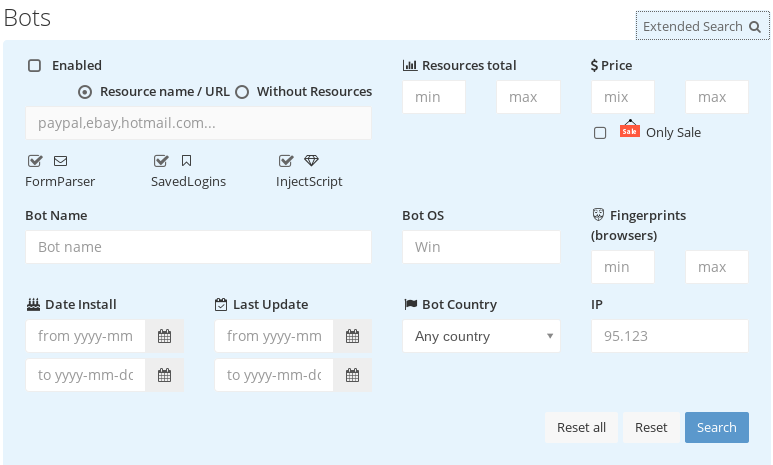}
\caption{View on the advanced search functionality.}
\label{fig:advancedsearch}
\end{figure}
\begin{figure*}[hbp!]
\includegraphics[width=1.8\columnwidth]{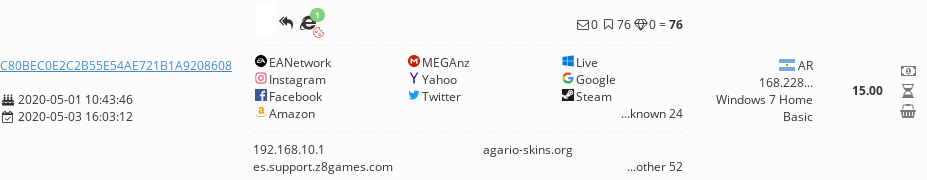}
\caption{Overview of a listed profile on \market.}
\label{fig:listing}
\end{figure*}

Figure~\ref{fig:advancedsearch} and Figure~\ref{fig:listing}
report screenshots from the market. 
Figure~\ref{fig:advancedsearch} depicts the search function of the market. Attackers can access a fine grained research tool that enables them to search for profiles with specific resource composition, number of available browser fingerprints, and other information.
In Figure~\ref{fig:listing}, an overview of the details for each profile is provided. On the left, from top to bottom, the name of the profile and the installation and update dates of the profile are listed. On top, in the center, the list of the available browsers (here only Microsoft Edge). The superimposed number indicates the number of fingerprints available for that specific browser and the superimposed icon whether cookies are available (green) or not (red). On the top-right is reported the number of resources available (here 76). %
In the center, an overview of the websites for which resources are present. On the right, details about the country, IP prefix and operating system are provided. Finally, to the right-most, there's the price expressed in USD and buttons to respectively buy, reserve or add the profile to the cart.

\section{Further data insights}

Table~\ref{table:coeff_sale} shows correlation coefficients for the logistic model $sale=\beta_0+\beta_1 Resources + \beta_2 Year + \beta_3 Cookies + \beta_4 Browsers$ (binary response variable). 
\begin{table}[hbp!]
\caption{Logistic regression for discounted profiles}
\label{table:coeff_sale}
\begin{tabular}{l c }
\toprule
$y=sale$                            & Model           \\
\midrule
$\beta_0$                   & $-0.09^*$       \\
                            & $(0.05)$        \\
\texttt{Resources}          & $0.00^{***}$    \\
                            & $(0.00)$        \\
\texttt{Year 2020}          & $-0.92^{***}$   \\
                            & $(0.04)$        \\
\texttt{Cookies}            & $0.00$          \\
                            & $(0.00)$        \\
\texttt{Browsers}           & $0.14^{***}$    \\
                            & $(0.02)$        \\
\midrule
$R^2$                       & $0.04$          \\
Num. obs.                   & $11683$         \\
\bottomrule
\multicolumn{2}{l}{\scriptsize{$^{***}p<0.001$, $^{**}p<0.01$, $^*p<0.05$}}
\end{tabular}
\end{table}
Whereas Resources, Year, and Browsers are significant predictors, the effect is very small with the unsurprising exception of Year, suggesting that profiles recently acquired are \textit{less} likely to be put on `sale'. The coefficient for $\beta_4$ shows a positive, albeit small, effect of the no. of browsers provided in a profile on the likelihood of profile sale. As indicated by the small $R^2$, we do not find a clear rationale explaining this effect.

\begin{table}[hbp!]
\caption{Autocorrelation matrix among categories of resources available for each bot.}
\label{tab:corr}
\begin{tabular}{p{0.097\textwidth}p{0.04\textwidth}p{0.038\textwidth}p{0.043\textwidth}p{0.038\textwidth}p{0.101\textwidth}}
\toprule
              & Crypto & Social & Services & Other & MoneyTransfer \\
              \midrule
Crypto        &        &        &          &       &               \\
Social        & 0.08   &        &          &       &               \\
Services      & 0.04   & 0.09   &          &       &               \\
Other         & 0.05   & 0.11   & 0.05     &       &               \\
MoneyTransfer & 0.16   & 0.26   & 0.10     & 0.17  &               \\
Commerce      & 0.12   & 0.28   & 0.09     & 0.17  & 0.42          \\
\bottomrule
\end{tabular}
\end{table}

Table~\ref{tab:corr} reports correlation coefficients between \resources\ types in our dataset. No high correlation is found, suggesting that no autocorrelation problem should affect the regression analysis provided in Section~\ref{sec:dataresources}.

\begin{table*}[hbp!]
\begin{center}
\caption{Statistical models for profiles sold at full price.}
\label{table:coefficientsfull}
\begin{tabular}{l c c c c c c c c c}
\toprule

 & Model 1 & Model 2 & Model 3 & Model 4 & Model 5 & Model 6 & Model 7 & Model 8 & Model 9 \\
\midrule
$\beta_0$                    & $8.71^{***}$  & $-12.11^{***}$ & $-11.64^{***}$ & $-2.08^{*}$  & $2.24^{**}$  & $-2.15^{**}$  & $-5.54^{***}$   & $-5.57^{***}$ & $-3.70^{***}$  \\
                             & $(0.08)$      & $(1.21)$       & $(1.13)$       & $(0.87)$     & $(0.80)$     & $(0.80)$      & $(0.82)$        & $(0.81)$      & $(0.63)$       \\
\texttt{\textit{Real} Fngrpr}& $1.06^{***}$  & $0.69^{***}$   & $0.80^{***}$   & $1.15^{***}$ & $1.30^{***}$ & $1.30^{***}$  & $1.30^{***}$    & $1.31^{***}$  & $1.11^{***}$   \\
                             & $(0.14)$      & $(0.16)$       & $(0.15)$       & $(0.11)$     & $(0.10)$     & $(0.10)$      & $(0.10)$        & $(0.10)$      & $(0.07)$       \\
log(GDP)                     &               & $2.29^{***}$   & $2.19^{***}$   & $0.87^{***}$ & $0.24^{**}$  & $0.49^{***}$  & $0.44^{***}$    & $0.46^{***}$  & $0.42^{***}$   \\
                             &               & $(0.12)$       & $(0.11)$       & $(0.09)$     & $(0.08)$     & $(0.08)$      & $(0.08)$        & $(0.08)$      & $(0.06)$       \\ 
\texttt{Crypto}              &               &                & $21.74^{***}$  & $15.19^{***}$& $14.15^{***}$& $13.87^{***}$ & $13.72^{***}$   & $13.62^{***}$ & $10.12^{***}$  \\
                             &               &                & $(0.65)$       & $(0.51)$     & $(0.46)$     & $(0.45)$      & $(0.44)$        & $(0.44)$      & $(0.34)$       \\
\texttt{Money Transfer}      &               &                &                & $12.30^{***}$& $9.91^{***}$ & $9.21^{***}$  & $9.07^{***}$    & $8.86^{***}$  & $6.20^{***}$   \\
                             &               &                &                & $(0.17)$     & $(0.17)$     & $(0.17)$      & $(0.16)$        & $(0.16)$      & $(0.13)$       \\ 
\texttt{Commerce}            &               &                &                &              & $5.94^{***}$ & $5.25^{***}$  & $5.27^{***}$    & $5.06^{***}$  & $3.22^{***}$   \\
                             &               &                &                &              & $(0.15)$     & $(0.15)$      & $(0.15)$        & $(0.15)$      & $(0.12)$       \\
\texttt{Social}              &               &                &                &              &              & $3.50^{***}$  & $3.52^{***}$    & $3.44^{***}$  & $1.68^{***}$   \\
                             &               &                &                &              &              & $(0.15)$      & $(0.15)$        & $(0.15)$      & $(0.12)$       \\
\texttt{Services}            &               &                &                &              &              &               & $4.08^{***}$    & $3.95^{***}$  & $2.31^{***}$   \\
                             &               &                &                &              &              &               & $(0.29)$        & $(0.29)$      & $(0.22)$       \\
\texttt{Other}               &               &                &                &              &              &               &                 & $4.22^{***}$  & $0.89^{***}$   \\
                             &               &                &                &              &              &               &                 & $(0.31)$      & $(0.24)$       \\
\texttt{Resources}           &               &                &                &              &              &               &                 &               & $0.10^{***}$   \\
                             &               &                &                &              &              &               &                 &               & $(0.00)$       \\

\midrule
R$^2$                        & <0.01         & 0.04           & 0.18           & 0.52         & 0.60         & 0.63          & 0.64            & 0.65          & 0.79          \\
Adj. R$^2$                   & <0.01         & 0.04           & 0.18           & 0.52         & 0.60         & 0.63          & 0.64            & 0.65          & 0.79          \\
Num. obs.                    & 7123          & 7123           & 7123           & 7123         & 7123         & 7123          & 7123            & 7123          & 7123          \\
\bottomrule
\multicolumn{10}{l}{\scriptsize{$^{***}p<0.001$, $^{**}p<0.01$, $^*p<0.05$}} \\
\end{tabular}
\end{center}
\end{table*}

\begin{table*}[hbp!]
\begin{center}
\caption{Statistical models for all profiles (sold at full price and on sale).}
\label{table:coefficientsfull_wsale}
\begin{tabular}{l c c c c c c c c c c}
\toprule

 & Model 1a & Model 2a & Model 3a & Model 4a & Model 5a & Model 6a & Model 7a & Model 8a & Model 9a & Model 10a \\
\midrule
$\beta_0$                    & $8.74^{***}$  & $-8.76^{***}$ & $-8.29^{***}$   & $-0.17$      & $3.52^{***}$  & $-0.29$      & $-3.49^{***}$   & $-3.46^{***}$ & $-2.26^{***}$ & $-2.10^{***}$ \\
                             & $(0.08)$      & $(0.85)$       & $(0.80)$       & $(0.63)$     & $(0.58)$     & $(0.58)$      & $(0.60)$        & $(0.59)$      & $(0.48)$      & $(0.44)$      \\
\texttt{\textit{Real} Fngrpr}& $1.07^{***}$  & $1.22^{***}$   & $1.29^{***}$   & $1.52^{***}$ & $1.66^{***}$ & $1.69^{***}$  & $1.69^{***}$    & $1.70^{***}$  & $1.49^{***}$  & $1.03^{***}$  \\
                             & $(0.14)$      & $(0.14)$       & $(0.13)$       & $(0.10)$     & $(0.09)$     & $(0.09)$      & $(0.09)$        & $(0.09)$      & $(0.07)$      & $(0.07)$      \\
log(GDP)                     &               & $1.77^{***}$   & $1.69^{***}$   & $0.58^{***}$ & $0.03$       & $0.25^{***}$  & $0.21^{***}$    & $0.21^{***}$  & $0.24^{***}$  & $0.36^{***}$  \\
                             &               & $(0.09)$       & $(0.08)$       & $(0.06)$     & $(0.06)$     & $(0.06)$      & $(0.06)$        & $(0.06)$      & $(0.05)$      & $(0.04)$      \\ 
\texttt{Crypto}              &               &                & $19.23^{***}$  & $13.76^{***}$& $12.72^{***}$& $12.48^{***}$ & $12.34^{***}$   & $12.25^{***}$ & $9.08^{***}$  & $8.91^{***}$  \\
                             &               &                & $(0.49)$       & $(0.39)$     & $(0.36)$     & $(0.35)$      & $(0.34)$        & $(0.34)$      & $(0.28)$      & $(0.25)$      \\
\texttt{Money Transfer}      &               &                &                & $10.95^{***}$& $8.82^{***}$ & $8.25^{***}$  & $8.12^{***}$    & $7.94^{***}$  & $5.58^{***}$  & $5.37^{***}$  \\
                             &               &                &                & $(0.13)$     & $(0.12)$     & $(0.12)$      & $(0.12)$        & $(0.12)$      & $(0.10)$      & $(0.09)$      \\ 
\texttt{Commerce}            &               &                &                &              & $5.21^{***}$ & $4.60^{***}$  & $4.60^{***}$    & $4.42^{***}$  & $2.71^{***}$  & $2.74^{***}$  \\
                             &               &                &                &              & $(0.11)$     & $(0.11)$      & $(0.11)$        & $(0.11)$      & $(0.09)$      & $(0.08)$      \\
\texttt{Social}              &               &                &                &              &              & $3.02^{***}$  & $3.08^{***}$    & $3.01^{***}$  & $1.42^{***}$  & $1.50^{***}$  \\
                             &               &                &                &              &              & $(0.11)$      & $(0.11)$        & $(0.11)$      & $(0.09)$      & $(0.08)$      \\
\texttt{Services}            &               &                &                &              &              &               & $3.83^{***}$    & $3.72^{***}$  & $2.13^{***}$  & $2.17^{***}$  \\
                             &               &                &                &              &              &               & $(0.21)$        & $(0.21)$      & $(0.17)$      & $(0.16)$      \\
\texttt{Other}               &               &                &                &              &              &               &                 & $3.54^{***}$  & $0.59^{**}$   & $0.69^{***}$  \\
                             &               &                &                &              &              &               &                 & $(0.22)$      & $(0.19)$      & $(0.17)$      \\
\texttt{Resources}           &               &                &                &              &              &               &                 &               & $0.10^{***}$  & $0.09^{***}$  \\
                             &               &                &                &              &              &               &                 &               & $(0.00)$      & $(0.00)$      \\
\texttt{Sale}                &               &                &                &              &              &               &                 &               &               & $-3.40^{***}$ \\
                             &               &                &                &              &              &               &                 &               &               & $(0.07)$      \\                             
                             
\midrule
R$^2$                        & <0.01         & 0.04           & 0.15           & 0.49         & 0.57         & 0.59          & 0.60            & 0.61          & 0.74          & 0.79          \\
Adj. R$^2$                   & <0.01         & 0.04           & 0.15           & 0.49         & 0.57         & 0.59          & 0.60            & 0.61          & 0.74          & 0.79          \\
Num. obs.                    & 11683         & 11683          & 11683          & 11683        & 11683        & 11683         & 11683           & 11683         & 11683         & 11683         \\
\bottomrule
\multicolumn{11}{l}{\scriptsize{$^{***}p<0.001$, $^{**}p<0.01$, $^*p<0.05$}} \\
\end{tabular}
\end{center}
\end{table*}

Table~\ref{table:coefficientsfull} reports all regression models on the expected (full) profile price. The main insight is that model coefficients are relatively stable as \resources\ are added in. When including bots on sale in the regression (Table~\ref{table:coefficientsfull_wsale}), coefficients appear relatively stable and in line with those reported in \autoref{table:coefficientsfull}, both in terms of trend and magnitude. An exception is $\log{(GDP)}$ in Model 5a, where the respective coefficient is not significant and drops in value when compared to Model 4a and Model 6a. This may suggest a correlation between $\log{(GDP)}$ and presence of \commerce\ resources for profiles on sale, that is not present or weaker for profiles at full price. %

\end{document}